\documentclass[a4paper,fleqn,usenatbib]{mnras}

\usepackage{newtxtext,newtxmath}

\usepackage[T1]{fontenc}
\usepackage{ae,aecompl}

\usepackage{amsfonts}
\usepackage{amsmath}
\usepackage[ngerman, british]{babel}
\usepackage{booktabs}
\usepackage{color}
\usepackage{enumitem}
\usepackage{graphicx}
\usepackage[utf8]{inputenc}
\usepackage{natbib}
\usepackage[]{tabularx}

\title[Debris discs around F~stars in $\beta$~Pic]{A $~75\%$ Occurrence Rate of Debris Discs around F stars in the $\beta$~Pic Moving Group
}

\author[N. Pawellek et al.]{
Nicole Pawellek,$^{1,2}$\thanks{E-mail: pawellek@ast.cam.ac.uk}
Mark Wyatt,$^{1}$
Luca Matr\`a,$^{3}$
Grant Kennedy, $^{4}$
Ben Yelverton, $^{1}$
\\
$^{1}$Institute of Astronomy, University of Cambridge, Madingley Road, Cambridge CB3 0HA, UK\\
$^{2}$Konkoly Observatory, Research Centre for Astronomy and Earth Sciences, Konkoly-Thege Mikl\'os \'ut 15-17, H-1121 Budapest, Hungary\\
$^{3}$ School of Physics, National University of Ireland Galway, University Road, Galway, Ireland\\
$^{4}$Department of Physics and Centre for Exoplanets and Habitability, University of Warwick, Gibbet Hill Road, Coventry CV4 7AL, UK\\
}

\date{Accepted XXX. Received YYY; in original form ZZZ}

\pubyear{2021}

\begin{document}
\label{firstpage}
\pagerange{\pageref{firstpage}--\pageref{lastpage}}
\maketitle

\begin{abstract}
Only 20\% of old field stars have detectable debris discs, leaving open the question of what disc, if any, is present around the remaining 80\%. Young moving groups allow to probe this population, since discs are expected to have been brighter early on. This paper considers the population of F~stars in the 23~Myr-old BPMG where we find that 9/12 targets possess discs. 
We also analyse archival ALMA data to derive radii for 4 of the discs, presenting the first image of the 63au radius disc of HD~164249.
Comparing the BPMG results to disc samples from $\sim45$~Myr and $\sim150$~Myr-old moving~groups, and to discs found around field stars, we find the disc incidence rate in young moving~groups is comparable to that of the BPMG and significantly higher than that of field~stars. 
The BPMG discs tend to be smaller than those around field~stars. However, this difference is not statistically significant due to the small number of targets. Yet, by analysing the fractional luminosity vs disc radius parameter space we find that the fractional luminosities in the populations considered drop by two orders of magnitude within the first 100~Myr. This is much faster than expected by collisional evolution, implying a decay equivalent to $1/\text{age}^2$. We attribute this depletion to embedded planets which would be around 170~$M_\text{earth}$ to cause a depletion on the appropriate timescale. However, we cannot rule out that different birth environments of nearby young clusters result in brighter debris discs than the progenitors of field~stars which likely formed in a more dense environment.

\end{abstract}




\section{Introduction}

After its protoplanetary disc has dispersed, a star is left with - if anything - a system of planets and debris belts. The dust in those debris belts is inferred to originate in the break-up of planetesimals at least kilometres in size \cite[e.g.,][]{wyatt-2008, krivov-2010, hughes-et-al-2018}, and is seen in far-infrared (FIR) surveys towards $\sim$20\% of nearby several Gyr-old stars \citep[e.g.,][]{eiroa-et-al-2013, sibthorpe-et-al-2018}, where a slightly higher detection rate is noted for earlier type stars \citep[e.g.,][]{su-et-al-2006, sibthorpe-et-al-2018}

FIR surveys of nearby stars also show that debris disc luminosities decrease with age in a manner explained by population models in which all stars are born with a debris belt that is depleted by collisions amongst the planetesimals \citep[][]{wyatt-et-al-2007, gaspar-et-al-2013}. This canonical model successfully explains the detection statistics (as a function of wavelength and stellar age), with the implication that all stars are born with a planetesimal belt of initial mass drawn from a log-normal distribution like that of protoplanetary discs, and concentrated at a radius drawn from a $n(r)\propto r^{-1.7}$ distribution in the range 1-1000~au \citep{sibthorpe-et-al-2018}.

However, while this model makes accurate predictions for the 20\% of Gyr-old stars with detectable discs, it is almost completely unconstrained for the 80\% of stars without detectable discs for which model predictions rely on the log-normal or power law assumptions about the underlying initial mass and radius distributions. For example, stars in the canonical model population without detectable discs are the 80\% with 1-10~au discs that are rapidly depleted and so never seen, whereas it would be equally valid to put these undetected discs at 30-100~au with very low initial mass. 
A further challenge comes from the inference that planetesimal belt radii depend on stellar luminosity. Belts imaged at millimetre wavelengths are larger around higher luminosity stars in a way that may be attributed to preferential formation at protoplanetary disc ice-lines \citep{matra-et-al-2018}, a possibility not currently included in the model.

It is inevitably challenging to determine the properties of the planetesimal belts of the 80\% of nearby stars without detectable dust. Our only hope is to probe these by studying stars that are young ($\ll100$~Myr when their discs are brightest) and nearby. Young nearby moving groups are ideal for sample selection, having also the benefit of being co-eval. Given the stellar luminosity dependence mentioned above, and that disc detection is maximised for higher luminosity stars, the optimal sample would include stars of similar early spectral type in the nearest youngest moving group. The number of A-type stars in nearby young moving groups for which the disc detection peaks is very limited while late-type stars are common. The best compromise between a high stellar luminosity and a reasonably large number of targets within the same moving group is given by F-type stars.

An example fulfilling the aforementioned requirements is the $\beta$~Pictoris moving group (BPMG) which contains stars of $\sim$23~Myr age \citep[][]{bell-et-al-2015}. 
Based on a survey of 30 BPMG stars of different spectral types, \cite{rebull-et-al-2008} found that more than 37\% of the targets show evidence for a circumstellar disc.
By considering the known F-type stars in the BPMG, \cite{churcher-et-al-2011} inferred a debris disc detection rate of 6/9 ($\sim67\%$). This is higher than the  $20\%$ seen for Gyr-old stars \citep[e.g.,][]{su-et-al-2006, marshall-et-al-2016, sibthorpe-et-al-2018} leading to the question why we get such a high frequency.
One explanation could be that during the timespan of two orders of magnitude in age ($\sim$20~Myr and $\sim$2~Gyr) a majority of discs is collisionally depleted so that we are not able to detect them anymore. Another possibility might be that the formation conditions of the BPMG differs significantly from these field stars leading to debris discs having atypical properties (i.e. unusually bright).

In the first part of this study we consider the population of F~star debris discs in the BPMG.
In \S~\ref{sec:sample} we revisit membership in the BPMG sample in the light of recent studies and note stellar multiplicity and planetary companions for the sample since those are possible influences on the occurrence of discs.
We investigate evidence for infrared excesses indicative of a surrounding debris disc in \S~\ref{sec:IRExcess}, then \S~\ref{sec:imaging} presents ALMA observations to determine the radii of the belts.
We use this spatial information to generate SED models including a size distribution of dust particles in \S~\ref{sec:SEDmodelling}.

In the second part of this study the properties of the BPMG disc population are compared with those of other nearby F-star populations in \S~\ref{sec:comparison} to identify similarities and differences between the samples.
In \S~\ref{sec:detectionrate} we analyse possible scenarios explaining the high detection rate of F~star debris discs in the BPMG, before concluding in \S~\ref{sec:conclusions}.

\section{Sample selection}
 \label{sec:sample}

 \subsection{Reassessing the BPMG sample of F stars}
The BPMG is one of the nearest moving groups. \cite{shkolnik-et-al-2017} identified 146 objects belonging to this group, where five stars are found to be A-type, eleven F-type, six G-type, 27 K-type and 97 M- and L-type. 
Using data from the \textit{Gaia} data release 2 \citep{gaia-collaboration-et-al-2018b}, several additional members of the BPMG were found by \cite{gagne-faherty-2018}. While the majority found in that study are M- and L-type, one F-type star and one A-type star could also be added to the sample given by \cite{shkolnik-et-al-2017}.
Thus, the sample of nine F-type members of the BPMG analysed by \cite{churcher-et-al-2011} is now increased to twelve by combining the samples of \cite{shkolnik-et-al-2017} and \cite{gagne-faherty-2018}. These twelve targets will be the basis of our analysis. 
All of them lie between 25 and 66~pc \citep{gaia-collaboration-et-al-2018b, bailer-jones-et-al-2018}, with stellar properties listed in Tab.~\ref{tab:sample}.

\begin{table*}
\caption{Stellar parameters of the sample of 12 F-type stars belonging to the $\beta$~Pic moving group.
\label{tab:sample}}
\tabcolsep 2pt
\centering
 \begin{tabular}{llcccc|llrrl|cc}
         &        &     &                  &                &        & \multicolumn{5}{c}{Multiplicity}                                                & \multicolumn{2}{c}{Planetary companion}\\
  HD     & HIP    & SpT & $L/L_\text{sun}$ & $T_\text{eff}$ & d      & Companion               & SpT       & \multicolumn{2}{c}{Separation}  & Ref     & Separation & Mass \\
         &        &     &                  & [K]            & [pc]   &                         &           & [arcsec] &[au]                  &         & [au]       & [$M_\text{Jup}$] \\
  \midrule
  203    & 560    & F2V & $4.26 \pm 0.03$  & $6830 \pm 30$  & 40.0   & \ldots                  & \ldots    & \ldots   & \ldots               & \ldots  & \ldots    & \ldots     \\
  14082A  & 10680  & F5V & $2.00 \pm 0.01$  & $6170 \pm 20$  & 39.8   & HD~14082B               & G2V       & 14       & 557                  & 1, 4    & \ldots    & \ldots     \\
  15115  & 11360  & F4V & $3.6  \pm 0.1$   & $6720 \pm 20$  & 49.0   & \ldots                  & \ldots    & \ldots   & \ldots               & \ldots  & \ldots    & \ldots     \\
  29391$^{a}$  & 21547  & F0V & $5.71 \pm 0.06$  & $7330 \pm 30$  & 29.8   & GJ~3305AB         & M1V       & 66       & 1957                 & 1       & 13        & 1-12 \\
  35850  & 25486  & F8V & $1.84 \pm 0.01$  & $6050 \pm 20$  & 26.9   & HD~35850B               & \ldots    & $7.8\times 10^{-4}$ & 0.021     & 6, 7    &\ldots     & \ldots     \\
  160305 & 86598  & F9V & $1.69 \pm 0.03$  & $6050 \pm 40$  & 65.7   & \ldots                  & \ldots    & \ldots   & \ldots               & \ldots  &\ldots     & \ldots     \\
  164249 & 88399  & F5V & $3.20 \pm 0.04$  & $6340 \pm 40$  & 49.6   & HD~164249B              & M2V       & 6.5      & 323                  & 1       &\ldots     & \ldots     \\
         &        &     &                  &                &        & 2MASS J18011138-5125594 & \ldots    & \ldots   & \ldots               & 3       &\ldots     & \ldots     \\
  173167 & -      & F5V & $2.4  \pm 0.1$   & $6270 \pm 90$  & 50.6   & TYC~9073-0762           & M1V       & 571      & 28894                & 1, 2    &\ldots     & \ldots     \\
  181327 & 95270  & F5V & $2.87 \pm 0.02$  & $6480 \pm 20$  & 48.2   & HD181296                & A0V+M7/8V & 416      & 20072                & 1       &\ldots     & \ldots     \\
  191089 & 99273  & F5V & $2.74 \pm 0.02$  & $6460 \pm 30$  & 50.1   & \ldots                  & \ldots    & \ldots   & \ldots               & \ldots  &\ldots     & \ldots     \\
  199143 & 103311 & F8V & $2.21 \pm 0.02$  & $5930 \pm 20$  & 45.7   & HD~199143B              & M2V       & 1.1      & 50                   & 4, 8    &\ldots     & \ldots     \\
         &        &     &                  &                &        & HD~358623               & K7        & 325.0    & 14764                & 1, 8    &\ldots     & \ldots     \\
  213429 & 111170 & F8V & $1.92 \pm 0.06$  & $5970 \pm 20$  & 25.5   & HD~213429B              & \ldots    & $\sim0.08^{b}$    & $\sim2^{b}$ & 1, 5    &\ldots     & \ldots     \\
 \bottomrule
 \end{tabular}
 
\noindent
{\em Notes:}
($^a$) HD~29391 is also known as 51~Eridani. The references for the planetary companion are given in Section~\ref{sec:planets}.
($^b$) The orbital period is 631~d. It is converted to a separation assuming the 
mass of the binary companion to be equal to that of the primary HD~213429 with $1.19M_\odot$. 

\noindent
{\em References for multiplicity:} 
[1] \cite{elliott-bayo-2016}, 
[2] \cite{moor-et-al-2013},
[3] \cite{gagne-et-al-2018b},
[4] \cite{mamajek-et-al-2014},
[5] \cite{kovaleva-et-al-2015},
[6] \cite{eker-et-al-2008},
[7] \cite{rodriguez-zuckerman-2012},
[8] \cite{tokovinin-1997}

\end{table*}

\subsection{Stellar multiplicity}
\label{sec:multiplicity}
Investigating the stellar multiplicity we found that our sample of F~stars possesses a 67\% fraction of multiple systems (8/12) including wide (separations $> 1000$~au) and very wide systems (separations $> 10000$~au). \cite{elliott-bayo-2016} studied the occurrence of such system configurations and found that the high fraction of multiples in the BPMG can be explained by the unfolding of primordial triple systems which was investigated by \cite{reipurth-mikkola-2012}. The term ``unfolding'' means that in triple systems, while born compact, one component is dynamically scattered into a very distant orbit within a few Myr. 
\cite{reipurth-mikkola-2012} showed that if the component scattered into a wide separation is of low mass while the close components are more massive, the triple system is likely to be unstable and disrupted on a short timescale into a massive binary system and a single low-mass star.
\cite{elliott-bayo-2016} find that the majority of the multiples' distant components in the BPMG are of low mass and therefore, the study predicts that these multiple systems should decay within 100~Myr. In our sample, the systems with low-mass binary companions are HD~29391, HD~164249, HD~173167 and HD~199143 for which we might expect a decay within the aforementioned time frame.

\subsection{Planetary companions}
\label{sec:planets}
In our sample of F~stars only the most luminous star HD~29391, also known as 51~Eridani, is known to possess a planetary companion \citep[e.g.,][]{macintosh-et-al-2015, nielsen-et-al-2019}. The system is located at a distance of 29.8~pc and forms a multiple stellar system with the M-type binary star GJ~3305AB \citep[e.g.,][]{janson-et-al-2014}. 
The companion 51~Eri~b was discovered by the Gemini Planet Imager Exoplanet Survey \citep[GPIES,][]{patience-et-al-2015, nielsen-et-al-2019} with a projected separation of 13~au. 
Depending on the formation model the estimated mass of the planet varies between 1\ldots2$M_\text{Jup}$ for a so-called ``hot start'' model \citep[][]{marley-et-al-2007, rajan-et-al-2017} and 2\ldots12$M_\text{Jup}$ for a ``cold start'' model \citep[][]{marley-et-al-2007, fortney-et-al-2008}.

\section{Assessing the sample for IR excess}
\label{sec:IRExcess}

\subsection{Modelling procedure}

We collected photometric data for all twelve targets in our sample from published catalogues, such as 2MASS \citep{cutri-et-al-2003}, the WISE All-Sky Release Catalog \citep{wright-et-al-2010}, the AKARI All-Sky Catalogue \citep{ishihara-et-al-2010}, the Spitzer Heritage Archive \citep[][]{carpenter-et-al-2008, lebouteiller-et-al-2011, chen-et-al-2014, sierchio-et-al-2014} and the Herschel Point Source Catalogue \citep{marton-et-al-2015}. These data allowed us to analyse the spectral energy distributions (SEDs) and therefore the occurrence of infrared emission in excess of that expected from the stellar photosphere. Mid- and far-infrared excesses are an indicator of the presence of a debris disc surrounding a host star. 

To find excesses we fit an SED model consisting of a star and a disc. 
We fit PHOENIX stellar photosphere models \citep{brott-hauschildt-2005} for each target using the stellar luminosity and the stellar temperature as model parameters. 
The resulting stellar luminosities and temperatures are listed in Tab.~\ref{tab:sample}. 
Knowing the stellar contribution to the mid- and far-infrared data we were able to derive the excess emission in the appropriate wavelength bands between 22 and 100$\mu$m taking into account the uncertainties of the photometry and the photospheric model \citep{yelverton-et-al-2019}. 
The results are given in Tab.~\ref{tab:IRexcess}.  

After subtracting the stellar emission the disc is fitted with a modified blackbody model \citep{backman-paresce-1993} for which the thermal emission of the dust is described as
\begin{equation}
 F_\nu \sim B_\nu(\lambda, T_\text{dust}) \left[H(\lambda_0 - \lambda) + H(\lambda - \lambda_0)\left(\frac{\lambda}{\lambda_0}\right)^{-\beta}\right],
\end{equation}
where $B_\nu$ is the Planck function and $H$ the Heaviside step function. The parameter $\lambda_0$ represents the characteristic wavelength while $\beta$ is the opacity index.
From this model we derive the dust temperature, $T_\text{BB}$, and the resulting blackbody radius of the disc, $R_\text{BB}$, as well as the fractional luminosity, $f_\text{d}$ (see Tab.~\ref{tab:SEDresults}). Here, $R_\text{BB}$ is the distance from the star that the temperature implies if the dust acted like blackbodies in equilibrium with the stellar radiation. 
In \S~\ref{sec:SEDmodelling} we apply a disc model including dust size distributions which do not exist in the framework of \cite{yelverton-et-al-2019}.   

The uncertainties of the fit parameters were inferred in the following way.
We start at the position of the minimum $\chi^2$ in parameter space, i.e. from the best fitting $f_\text{d}$ and $R_\text{BB}$. 
A set of new parameter values is randomly generated from which we calculate the SED. This leads to a new $\chi^2$ value which is compared to the former minimum value. The $\chi^2$ parameter estimates how likely the set of parameter values fits the SED. If the probability is larger than a certain threshold value the set is saved. 
In the end, it is counted how often the code reaches a certain set of $f_\text{d}$ and $R_\text{BB}$. The closer the parameters get to the best fitting values the higher the probability. 
The resulting distribution in parameter space represents an estimate for the probability distribution of the parameters and thus allows us to calculate the confidence levels for the parameters assuming that the values follow a normal distribution \citep[simulated annealing; e.g.,][]{pawellek-2017}.

\subsection{Stars with IR excess}

We identified nine out of twelve stars ($75\%$) that show infrared excess and therefore suggest the presence of a debris disc. 
Since we cannot draw strong conclusions on HD~173167 (see \S~\ref{sec:targetnotes}) we might even say that nine out of eleven systems ($\sim82\%$) 
possess debris discs.
A comparable, high detection rate of 6/9 F~stars for the BPMG was noted in \cite{churcher-et-al-2011}. As noted in the introduction, this is in contrast to the results of studies which find a typical occurrence rate for debris discs of $\sim$20\% around FGK-stars for volume limited samples with older mean ages around Gyr \citep[e.g.,][]{su-et-al-2006, eiroa-et-al-2013, chen-et-al-2014, marshall-et-al-2016, sibthorpe-et-al-2018}.

The fractional luminosities of the excess emission lie between $1.2\times 10^{-5}$ and $4.1\times10^{-3}$, which are typical values for debris discs \citep[e.g.,][]{eiroa-et-al-2013, chen-et-al-2014, holland-et-al-2017, sibthorpe-et-al-2018}. 
The inferred blackbody temperatures lie between 51 and 600~K corresponding to blackbody radii between 0.3 and 52~au. 
 
\cite{pawellek-krivov-2015} found a relation between the ratio of the spatially resolved disc radius seen at FIR wavelengths to blackbody radius and the stellar luminosity of the form 
\begin{equation}
 \frac{R_\text{FIR}}{R_\text{BB}} = A \left(\frac{L}{L_\text{sun}}\right)^B.
 \label{eq:trueradius}
\end{equation}
The disc radius seen at FIR wavelengths in this relation is that inferred from resolved \textit{Herschel}/PACS imaging, and the blackbody radius that of a fit to the spectrum that is comparable with the modified blackbody fit used here. 
We use the updated values of $A$ and $B$ from \cite{pawellek-2017} with $A = 6.49\pm0.86$ and $B = -0.37\pm0.05$ assuming pure astronomical silicate \citep{draine-2003} for the dust material. 

Estimates of the FIR radii of the discs using eq.~(\ref{eq:trueradius}) give values between 1.5 and 215~au which are $\sim$4 times larger than $R_\text{BB}$. 
In \S\ref{sec:trueradius} we compare those estimates to the observed disc radii of the spatially resolved discs. 

\subsection{Notes for particular targets}
\label{sec:targetnotes}
HD~14082A: For HD~14082A all WISE bands \citep[3.4, 4.6, 12 and 22$\mu$m, see WISE All-Sky Catalog,][]{wright-et-al-2010} and Spitzer/MIPS \citep[24$\mu$m,][]{chen-et-al-2014} exhibit significant excess emission, but no excess was found with Spitzer/MIPS (70$\mu$m) or Herschel/PACS.
The star forms a binary system \citep{mamajek-et-al-2014, elliott-bayo-2016} with its companion (HD~14082B) known to exhibit IR excess in the mid- and far-infrared \citep[e.g.,][]{riviere-marichalar-et-al-2014}.  
After checking the WISE and MIPS data we found the photometry to be confused in all bands since those instruments were not able to differentiate between the two stellar components. Thus, we assume no significant excess emission for HD~14082A while the excess found around HD~14082B is real.

HD~29391: The star HD~29391 (51~Eri) shows significant excess at MIPS24, MIPS70 and PACS100 providing a good constraint on the disc as noted previously  \citep{riviere-marichalar-et-al-2014}. The target possesses the only planetary companion detected in our sample (see \S\ref{sec:planets}). The planet's separation is $\sim13$~au. 
With the disc's $R_\text{BB} = 9\pm2$~au and an estimated FIR radius of $R_\text{FIR} = 30.7\pm13.4$~au we assume the planet to be located closer to the star than the planetesimal belt.

HD~173167: The star HD~173167 only possesses photometric data up to 22$\mu$m so that we cannot draw any conclusions about a possible far-infrared excess. However, the mid-infrared data do not reveal significant excess. 

HD~199143: Considering HD~199143, there are mid-infrared data available as well as data from \textit{Herschel}/PACS. The excess emission is significant only for the MIPS24 band, but WISE data also show a marginal detection of excess emission at 22$\mu$m. Despite the presence of a close binary companion we could rule out confusion issues since the companion is several orders of magnitude fainter than the primary. 
Therefore, we assume that we detect emission from hot dust close to the star. In our sample this is the only target with a close-in disc, with a dust temperature of 600~K and a blackbody radius of 0.3~au. The FIR radius is estimated to be 1.5~au. 

While Tab.~\ref{tab:IRexcess} gives the significance of the excess at 24$\mu$m as $7\sigma$ this might be overestimated because of the SED fit being to both star and disc. Fitting the star without including the disc component results in a 24$\mu$m excess of $3\sigma$. Thus the excess is real, albeit at low significance.

\begin{table*}
\caption{IR excesses.
\label{tab:IRexcess}}
\tabcolsep 1pt
\centering
 \begin{tabular}{r|rcrcr|rcrcr|rcrcr|rcrcr|rcrcr}
       & \multicolumn{5}{c}{WISE22}  & \multicolumn{5}{c}{MIPS24} & \multicolumn{5}{c}{MIPS70} & \multicolumn{5}{c}{PACS70} & \multicolumn{5}{c}{PACS100}            \\
HD     & \multicolumn{3}{c}{$F_\nu$} & $F_{\nu,\star}$      & Excess   & \multicolumn{3}{c}{$F_\nu$} & $F_{\nu,\star}$    & Excess   & \multicolumn{3}{c}{$F_\nu$}    & $F_{\nu,\star}$   & Excess   & \multicolumn{3}{c}{$F_\nu$}   & $F_{\nu,\star}$     & Excess   & \multicolumn{3}{c}{$F_\nu$}   & $F_{\nu,\star}$    & Excess   \\
       & \multicolumn{3}{c}{[mJy]}   & [mJy]     & $\sigma$ & \multicolumn{3}{c}{[mJy]}  & [mJy]     & $\sigma$ & \multicolumn{3}{c}{[mJy]}   & [mJy]      & $\sigma$ & \multicolumn{3}{c}{[mJy]}    & [mJy]      & $\sigma$ & \multicolumn{3}{c}{[mJy]}    & [mJy]     & $\sigma$ \\
\midrule
203    & 126.5 & $\pm$ & 7.2 & 64.2 & 8.6      & 120.5 &$\pm$& 2.4          & 55.9 & 27     & 61.0  & $\pm$ & 10.4       & 6.1 & 5.3    & 71.5  & $\pm$ & 4.4        & 6.4 & 15     & 41.3  & $\pm$ & 2.5        & 3.2 & 15     \\
14082A  & 64.1  & $\pm$ & 3.8 & 41.3 & 6.0$^{*}$& 39.9  &$\pm$& 0.8          & 36.0 & 4.9$^{*}$& \multicolumn{3}{c}{\ldots} & 3.9 & \ldots & \multicolumn{3}{c}{$<13$}  & 4.1 & \ldots & \multicolumn{3}{c}{$<8$}   & 2.0 & \ldots \\
15115  & 63.1  & $\pm$ & 3.8 & 39.0 & 6.3      & 58.3  &$\pm$& 2.3          & 33.9 & 11     & 451.9 & $\pm$ & 32.6       & 3.7 & 14     & 463.0 & $\pm$ & 14.5       & 3.9 & 32     & \multicolumn{3}{c}{\ldots} & 1.9 & \ldots \\
29391  & 141.8 & $\pm$ & 8.0 & 123  & 2.3      & 129.7 &$\pm$& 2.6          & 107  & 8.8    & 23.0  & $\pm$ & 0.92       &11.5 & 12     & 21.8  & $\pm$ & 3.8        &12.2 & 2.5    & 19.0  & $\pm$ & 3.0        & 6.0 & 4.3    \\
35850  & 96.9  & $\pm$ & 5.7 & 87.5 & 1.6      & 83.5  &$\pm$& 3.4          & 76.2 & 2.1    & 40.3  & $\pm$ & 9.20       & 8.3 & 3.5    & \multicolumn{3}{c}{\ldots} & 8.8 & \ldots & 46.1  & $\pm$ & 2.6        & 4.3 & 16     \\
160305 & 18.5  & $\pm$ & 1.7 & 13.2 & 3.1      & \multicolumn{3}{c}{\ldots} & 11.5 & \ldots & \multicolumn{3}{c}{\ldots} & 1.3 & \ldots & 31.9  & $\pm$ & 1.6        & 1.3 & 18     & \multicolumn{3}{c}{\ldots} & 0.65& \ldots \\
164249 & 85.4  & $\pm$ & 5.1 & 39.2 & 9.0      & 77.4  &$\pm$& 1.6          & 34.1 & 28     & 624.1 & $\pm$ & 62.4       & 4.1 & 10     & \multicolumn{3}{c}{\ldots} & 3.9 & \ldots & 513.0 & $\pm$ & 17.7       & 1.9 & 29     \\
173167 & 33.9  & $\pm$ & 2.2 & 29.3 & 2.0      & \multicolumn{3}{c}{\ldots} & 25.5 & \ldots & \multicolumn{3}{c}{\ldots} & 3.7 & \ldots & \multicolumn{3}{c}{\ldots} & 2.9 & \ldots & \multicolumn{3}{c}{\ldots} & 1.4 & \ldots \\
181327 & 212.2 & $\pm$ & 12.1& 34.6 & 15       & 205.4 &$\pm$& 4.1          & 30.1 & 43     & 1468  & $\pm$ & 249        & 3.3 & 5.9    & \multicolumn{3}{c}{\ldots} & 3.5 & \ldots & 1463  & $\pm$ & 47         & 1.7 & 31     \\
191089 & 192.8 & $\pm$ & 10.9& 30.9 & 15       & 187.5 &$\pm$& 3.8          & 26.9 & 43     & 544.3 & $\pm$ & 50.6       & 2.9 & 11     & \multicolumn{3}{c}{\ldots} & 3.1 & \ldots & 422.6 & $\pm$ & 13.5       & 1.5 & 31     \\
199143 & 45.2  & $\pm$ & 2.9 & 37.8 & 2.5      & 38.8  &$\pm$& 0.9          & 32.9 & 7.0    & \multicolumn{3}{c}{$<9$}   & 3.6 & \ldots & 4.89  & $\pm$ & 1.37       & 3.8 & 0.80   & \multicolumn{3}{c}{\ldots} & 1.9 & \ldots \\
213429 & 107.3 & $\pm$ & 6.3 & 105  & 0.40     & 93.1  &$\pm$& 1.9          & 91.2 & 1.0    & 22.2  & $\pm$ & 4.1        &10.0 & 2.9    &\multicolumn{3}{c}{\ldots}  &10.5 & \ldots & \multicolumn{3}{c}{\ldots} & 5.2 & \ldots \\
 \bottomrule

 \end{tabular}
 
 \noindent
{\em Notes:}
Errors include both statistical and systematic uncertainties. 
WISE data from WISE All-Sky Catalog \citep{wright-et-al-2010}, MIPS24 and MIPS70 from Spitzer Heritage Archive \citep{carpenter-et-al-2008, chen-et-al-2014, sierchio-et-al-2014}, PACS70 and PACS100 from Herschel Science Archive \citep{eiroa-et-al-2013}. Upper limits stem from \cite{riviere-marichalar-et-al-2014}.
The thermal emission caused by the dust material surrounding the star is given as excess from the stellar photosphere in units of $\sigma$ and is considered to be significant if it reaches a value larger than 3.
$^{(*)}$ The WISE and MIPS data of HD~14082A were found to be confused and thus, are not taken into account when checking the presence of IR excess.
\end{table*}

\begin{table*}
 \caption{SED fitting results.
 \label{tab:SEDresults}}
\tabcolsep 1pt
\centering
 \begin{tabular}{rc|ccccccccccc|ccccccccccccccccc}
HD & Resolved & \multicolumn{11}{c}{Modified blackbody model}                                                                              & \multicolumn{17}{c}{Grain size distribution model} \\
    &          & $f_\text{d}$  & \multicolumn{3}{c}{$T_\text{bb}$} & \multicolumn{3}{c}{$R_\text{bb}$} & \multicolumn{3}{c}{$R_\text{est, sub-mm}$} && \multicolumn{3}{c}{$R_\text{sub-mm}$} & $s_\text{blow}$ & \multicolumn{3}{c}{$s_\text{min}$} & \multicolumn{3}{c}{$s_\text{min}/s_\text{blow}$} & \multicolumn{3}{c}{$q_\text{SED}$} & \multicolumn{3}{c}{$T_\text{dust}$} & $f_\text{d}$   \\
	&          & [$10^{-5}$]   & \multicolumn{3}{c}{[K]}           & \multicolumn{3}{c}{[au]}          & \multicolumn{3}{c}{[au]}           && \multicolumn{3}{c}{[au]}            & [$\mu$m]        & \multicolumn{3}{c}{[$\mu$m]}       & \multicolumn{3}{c}{}                             & \multicolumn{3}{c}{}               & \multicolumn{3}{c}{[K]}             &  [$10^{-5}$]   \\              \midrule

203    & No     & $15$  & 134 &$\pm$& 4             & 8.3 &$\pm$& 1.5          & 30  &$\pm$& 8             &&\multicolumn{3}{c}{\ldots} & 1.18 &\multicolumn{3}{c}{\ldots}&\multicolumn{3}{c}{\ldots}& \multicolumn{3}{c}{\ldots}&\multicolumn{3}{c}{\ldots}&\ldots\\
14082A & \ldots &\ldots &\multicolumn{3}{c}{\ldots} &\multicolumn{3}{c}{\ldots} &\multicolumn{3}{c}{\ldots} &&\multicolumn{3}{c}{\ldots} & 0.87 &\multicolumn{3}{c}{\ldots}&\multicolumn{3}{c}{\ldots}& \multicolumn{3}{c}{\ldots}&\multicolumn{3}{c}{\ldots}&\ldots\\
15115  & Yes    &$51$   &  62 &$\pm$&  1            & 38  &$\pm$& 7             &129  &$\pm$& 27            && 93 &$\pm$& 21             & 0.91 & 4.6 &$\pm$& 0.2 & 5.1 &$\pm$& 0.2 & 3.84 &$\pm$& 0.06 & 61.4 &$\pm$& 1.9 & $60.0$\\ 
29391  & No     &$0.5$  & 101 &$\pm$& 20            &  18  &$\pm$& 5             & 21  &$\pm$& 9             &&\multicolumn{3}{c}{\ldots} & 1.81 &\multicolumn{3}{c}{\ldots}&\multicolumn{3}{c}{\ldots}& \multicolumn{3}{c}{\ldots}&\multicolumn{3}{c}{\ldots}&\ldots\\
35850  & No     &$4.0$  &  74 &$\pm$&  4            & 19  &$\pm$& 8             & 54  &$\pm$& 6             &&\multicolumn{3}{c}{\ldots} & 0.51 &\multicolumn{3}{c}{\ldots}&\multicolumn{3}{c}{\ldots}& \multicolumn{3}{c}{\ldots}&\multicolumn{3}{c}{\ldots}&\ldots\\
160305 & Yes    &$14$   &  56 &$\pm$& 10            & 32  &$\pm$& 9             & 71  &$\pm$& 23            && 88 &$\pm$&  2             & 0.67 & 0.6 &$\pm$& 0.6 & 0.9 &$\pm$& 0.8 & \multicolumn{3}{c}{3.5$^a$}& 77.4 &$\pm$& 14.5 & 22.4 \\
164249 & Yes    &$88$   &  60 &$\pm$&  1            & 39  &$\pm$& 8             & 93  &$\pm$& 15            && 63 &$\pm$& 24             & 0.89 & 2.8 &$\pm$& 0.1 & 3.2 &$\pm$& 0.2 & 3.73 &$\pm$& 0.05 & 77.4 &$\pm$& 2.6 & $94.2$\\ 
173167 & \ldots &\ldots &\multicolumn{3}{c}{\ldots} &\multicolumn{3}{c}{\ldots} &\multicolumn{3}{c}{\ldots} &&\multicolumn{3}{c}{\ldots} & 0.85 &\multicolumn{3}{c}{\ldots}&\multicolumn{3}{c}{\ldots}& \multicolumn{3}{c}{\ldots}&\multicolumn{3}{c}{\ldots}&\ldots\\
181327 & Yes    &$264$  &   78&$\pm$&  1            & 22  &$\pm$& 4             &125  &$\pm$& 22            && 81 &$\pm$& 16             & 1.02 & 1.1 &$\pm$& 0.2 & 1.1 &$\pm$& 0.2 & 3.45 &$\pm$& 0.05 & 61.4 &$\pm$& 4.7 & $227$\\ 
191089 & Yes    &$151$  & 92&$\pm$&1            & 15&$\pm$& 3           & 37  &$\pm$& 4             && 45 &$\pm$& 16             & 0.98 & 1.2 &$\pm$& 0.3 & 1.2 &$\pm$& 0.3 & 3.43 &$\pm$& 0.08 & 83.6 &$\pm$& 4.9 & $118$\\ 
199143 & No     & $47$  & 1000 &$\pm$&100            &  0.2&$\pm$& 0.1           & 0.8 &$\pm$& 0.3           &&\multicolumn{3}{c}{\ldots} & 0.74 &\multicolumn{3}{c}{\ldots}&\multicolumn{3}{c}{\ldots}& \multicolumn{3}{c}{\ldots}&\multicolumn{3}{c}{\ldots}&\ldots\\
213429 & \ldots &\ldots &\multicolumn{3}{c}{\ldots} &\multicolumn{3}{c}{\ldots} &\multicolumn{3}{c}{\ldots} &&\multicolumn{3}{c}{\ldots} & 0.74 &\multicolumn{3}{c}{\ldots}&\multicolumn{3}{c}{\ldots}& \multicolumn{3}{c}{\ldots}&\multicolumn{3}{c}{\ldots}&\ldots\\

 \end{tabular}

\noindent
{\em Notes:}
The blow-out limit is calculated assuming Mie theory and pure astronomical silicate \citep[][]{draine-2003} with a bulk density of $3.3$ g/cm$^3$. 
The estimated disc radius seen at sub-mm wavelengths, $R_\text{est, sub-mm}$, is calculated by eq.~(\ref{eq:trueradius}) using the parameters inferred in this work. A grain size distribution fit is done if the disc is spatially resolved.
$^a$ The parameter $q_\text{SED}$ was fixed due to a lack of photometric data in the far-infrared. Only HD~15115 shows evidence for a warm disc component.
\end{table*}

\begin{figure*}
\centering
\includegraphics[width=0.9\textwidth]{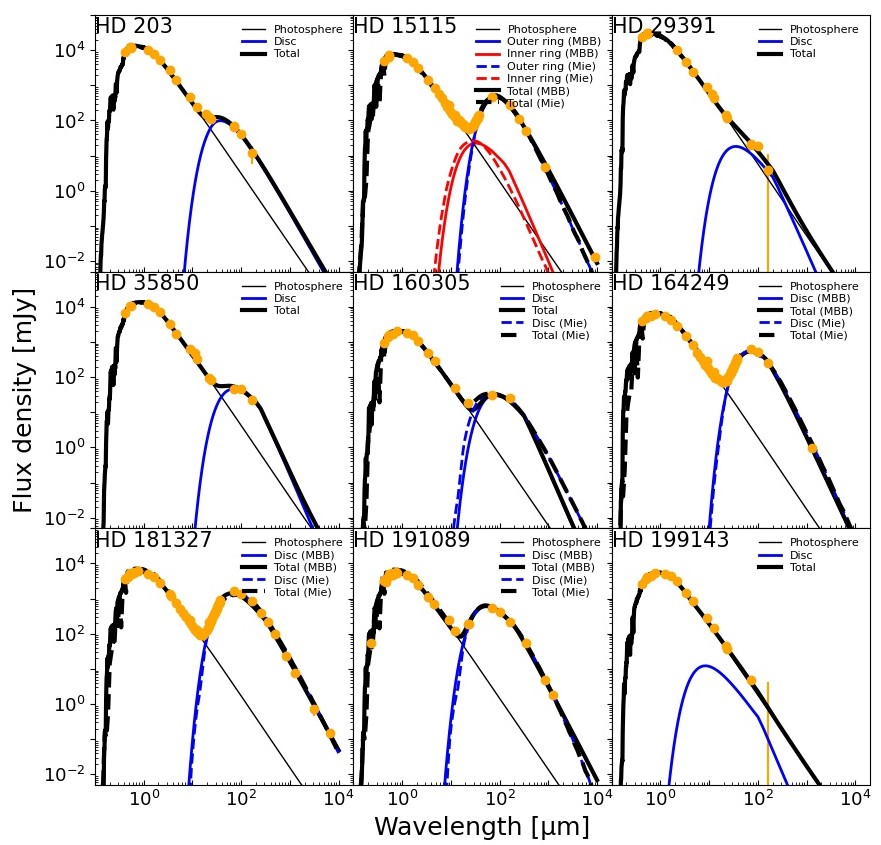}
 \caption{SEDs for the debris discs detected around F stars in the BPMG. Solid lines show the modified blackbody fit. For spatially resolved targets, dashed lines show the size distribution fit using Mie theory. Blue lines represent the outer ring, red lines the inner ring (if present).
For the SED of HD~15115 both disc components were fitted with a modified blackbody model (solid line) and a size distribution model (dashed line).
 \label{fig:SEDs}}
\end{figure*}

\subsection{IR excess in multiple systems}

We found that all four stars in our sample without a known stellar companion possess a debris disc (HD~203, HD~15115, HD~160305 and HD~191089). Furthermore, three out of the five systems with companions at projected separations larger than 135~au (HD~14082A, HD~29391, HD~164249, HD~173167 and HD~181327) harbour a disc as well.  
Two systems have companions at projected separations below 25~au where one shows evidence of debris. 
(HD~35850 has debris and a companion at a distance of 0.021~au, while
HD~213429 has no debris and a companion with an estimated separation of $\sim$2~au). 
Only HD~199143 has a stellar companion at an intermediate separation of $\sim$50~au \citep[in addition to a wide separation component at $\sim15000$~au, ][]{tokovinin-1997, mamajek-et-al-2014}.
Significant mid-infrared excess (see \S~\ref{sec:targetnotes}) hints at the presence of a close-in debris disc with $R_\text{BB} = 0.3$~au.

These results are broadly consistent with those of \cite{yelverton-et-al-2019}.
That study analysed a sample of 341 multiple systems and found that for binary stars with separations between $\sim$25 and 135~au no discs could be detected. Since these values are comparable to typical debris disc radii \citep[e.g.,][]{booth-et-al-2013, pawellek-et-al-2014, matra-et-al-2018} it was suggested that the binaries are clearing the primordial circumstellar or circumbinary material via dynamical perturbation. While the detection rates for separations larger than 135~au were found to be similar to the rates for single stars (at $\sim$20\%), only $\sim8$\% of binary systems with separations below 25~au showed evidence for debris.  

Thus, considering the three aforementioned targets (HD~35850, HD~199143, HD~213429), we would expect a lower number of disc detections for these systems, but as they are only three in number we cannot draw any conclusions about detection rates.

\section{Disc imaging}
\label{sec:imaging}

Different observational wavelengths are sensitive to different sizes of dust grains. While the emission seen by (sub-)mm telescopes such as ALMA is expected to be dominated by thermal emission from mm-sized particles, near-infrared instruments such as VLT/SPHERE \citep{beuzit-et-al-2019} are expected to trace scattered light from micron-sized grains. 
Particles as small as micron-size are significantly affected by radiation pressure and other transport processes \citep[e.g.,][]{burns-et-al-1979} so that their distribution is expected to extend far beyond their birth environment. In contrast the large mm-sized grains are expected to stay close to the parent belt. 
Hence, the disc size inferred in sub-mm observations is the best tracer of the location of a system's planetesimal belt, which might differ from the radial extent seen in near-infrared scattered light. Nevertheless, such short wavelength observations can also be modelled to infer the planetesimal belt location, and comparing the disc structure seen at different wavelengths provides information on the physical mechanisms shaping debris discs.

\subsection{ALMA reduction procedure}

Atacama Large Millimeter/submillimeter Array (ALMA) observations of six out of the twelve stars in our sample were retrieved from the ALMA archive. 
Three of the analysed datasets have already been presented in literature work \citep[HD~15115, HD~29391, HD~191089,][]{macgregor-et-al-2019, perez-et-al-2019, kral-et-al-2020}, but are re-analysed here to maintain consistency across the sample. 
We also present the first ALMA image of HD~164249 and new images of the disc around HD~181327. For the latter target another dataset was published by \cite{marino-et-al-2016}, but due to their lower resolution we do not use those data here.
In addition to that we present new constraints for HD~14082A for which dust emission was not detected (as was also the case for HD~29391).

The targets were observed as single pointings with the ALMA 12~m array within the context of a variety of projects, over different ALMA Cycles, leading to inhomogeneous sensitivities, resolutions, and wavelengths (see Table \ref{tab:obslog}). For each target, and each observing date, we carried out standard calibration steps to obtain calibrated visibility datasets; we used the same CASA and pipeline version as used in the original reduction delivered by the ALMA observatory.

Later processing was carried out homogeneously in CASA v5.4.0. If available, for each target, we concatenated multiple datasets at similar frequencies to obtain a final combined visibility dataset. We also averaged in time (to 30s integrations) and frequency (to 2~GHz-wide channels) to reduce the size of each dataset and speed up imaging and modelling.

We then carried out continuum imaging using the CLEAN algorithm implemented through the \textit{tclean} CASA task. We used multiscale deconvolution \citep{cornwell-2008} in multi-frequency synthesis mode, adapting the choice of visibility weighting and/or tapering schemes to achieve a good trade-off between image sensitivity and resolution. The weighting choices, RMS noise levels (measured in image regions free of source emission), and synthesised beam sizes achieved are listed in Table \ref{tab:obslog}.

\begin{table*}
\caption{BPMG F~stars ALMA observations Summary 
\label{tab:obslog}}
\begin{tabular}{cccccccc}
Target & Date & Project ID & Band & Continuum & Continuum & Cont. Image & Original Reference \\
       &      &            &      & RMS       & Beam size & weighting   & for Dataset        \\
       & dd-mm-yyyy &  & & $\mu$Jy beam$^{-1}$ &   &  & \\
\midrule

HD14082A
 & 31-08-2015 & 2013.1.01147 & 6 & 150$^{a}$ & 1.8$\arcsec$ $\times$1.6$\arcsec$ & Natural & This work \\
HD15115
 & 01-01-2016 & 2015.1.00633 & 6 & 15 & 0.6$\arcsec$ $\times$0.6$\arcsec$ & Briggs 0.5 & \citet{macgregor-et-al-2019} \\
 & 09-06-2016 & 2015.1.00633 & 6 &  &  &  & \citet{macgregor-et-al-2019} \\
HD29391
 & 13-10-2016 & 2016.1.00358 & 6 & 23 & 0.2$\arcsec$ $\times$0.2$\arcsec$ & Natural & \cite{perez-et-al-2019} \\
HD164249
 & 10-03-2014 & 2012.1.00437 & 6 & 45 & 1.1$\arcsec$ $\times$1.0$\arcsec$ & 0.7$\arcsec$ Taper & This work \\
 & 11-08-2015 & 2013.1.01147 & 6 & &  & & This work \\
HD181327
 & 25-07-2016 & 2015.1.00032 & 7 & 27 & 0.2$\arcsec$ $\times$0.2$\arcsec$ & Natural & This work \\
HD191089
 & 23-03-2014 & 2012.1.00437 & 6 & 12 & 0.9$\arcsec$ $\times$0.6$\arcsec$ & Briggs 0.0 & This work \\
 & 30-05-2018 & 2017.1.00704 & 6 & & & & \citet{kral-et-al-2020} \\
 & 14-09-2018 & 2017.1.00200 & 6 & & & & Matr\`a et al. (in prep.) \\
\bottomrule
\end{tabular}

\noindent
{\em Notes:}
$^{a}$ At field center. HD14082A is however $\sim14\arcsec$ from field center, where the primary beam level drops to 46\%. The sensitivity per beam at that location would be $326$ $\mu$Jy beam$^{-1}$.
RMS noise levels, beam sizes and weightings of multiple datasets refer to imaging of the joint datasets.

\end{table*}

\begin{figure*}
 \includegraphics[width=\textwidth]{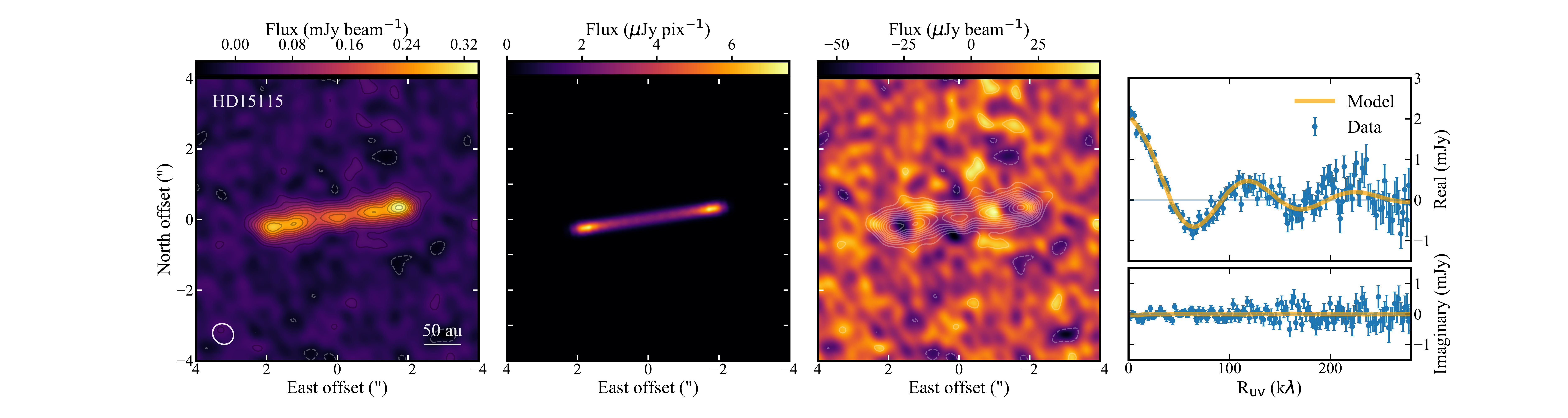}
 \includegraphics[width=\textwidth]{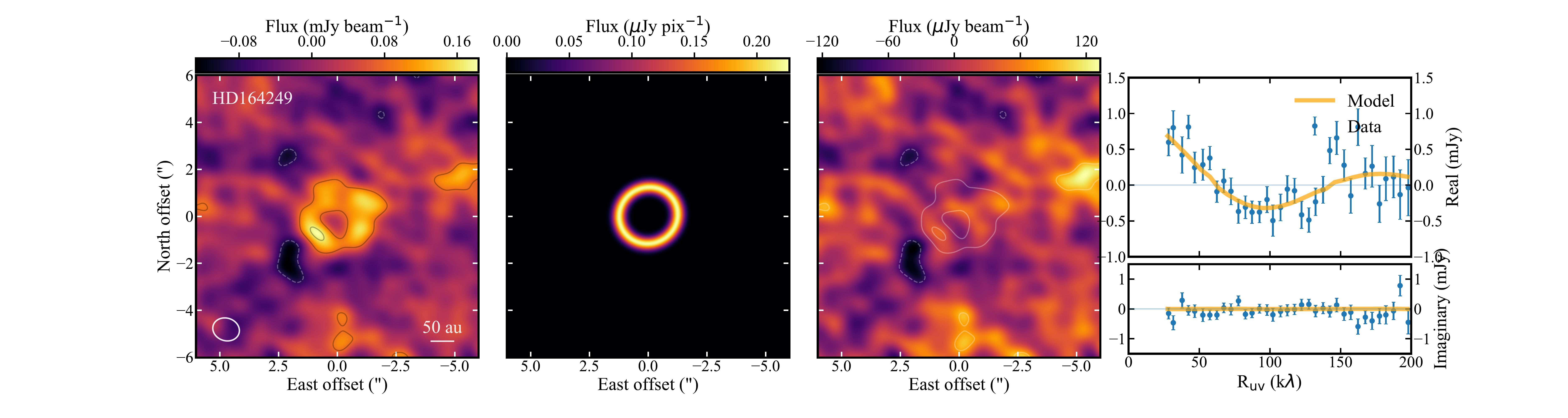}
 \includegraphics[width=\textwidth]{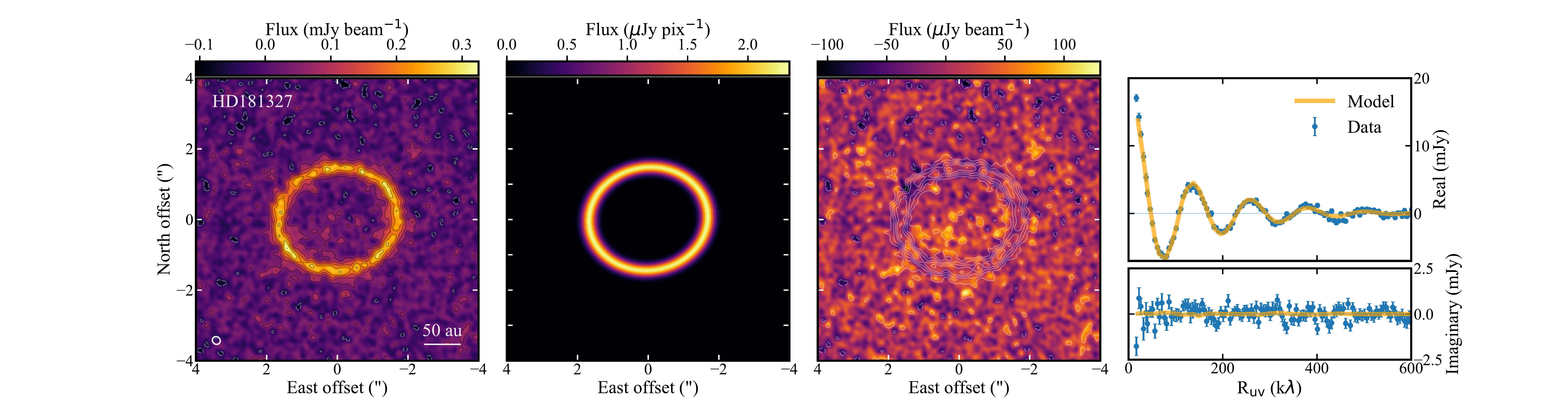}
 \includegraphics[width=\textwidth]{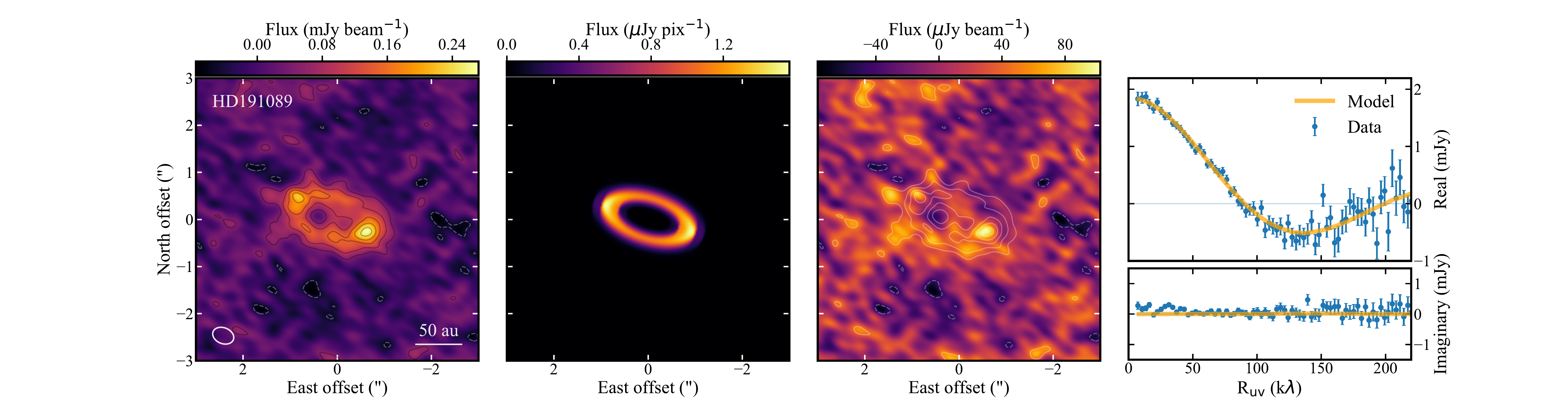}
 \caption{ALMA models for four resolved debris discs. From top to bottom: HD~15115, HD~164249, HD~181327, HD~191089.
 \label{fig:ALMA_Models}}
\end{figure*}

Discs are detected and resolved around four out of the six BPMG F~stars with existing ALMA observations. Fig.~\ref{fig:ALMA_Models} shows the ALMA images for the resolved discs, as well as the best-fit models, residuals, and deprojected visibilities. 
No detection was achieved near the location of HD~14082A and HD~29391 in the respective images. 
We conservatively derive 3$\sigma$ upper limits of $<5.8$ and $<3.5$~mJy for the flux density of the two belts, respectively, by spatially integrating emission within a 5$\arcsec$ radius circular region centered on the expected stellar location. 
The high values for the upper limits are caused by the relatively small beam used for the observation of HD~29391 and the fact that HD~14082A is significantly offset from the phase center, increasing the already high RMS of that observation. For both targets no (sub-)mm observations were reported in the literature before. 

For the visibility modelling, we follow the method described e.g. in \citet{matra-et-al-2019b}, using RADMC-3D\footnote{\url{http://www.ita.uni-heidelberg.de/~dullemond/software/radmc-3d/}} to calculate model images from a given density distribution, which we here assume to be a radial and vertical Gaussian described by
\begin{equation}
\rho=\Sigma_0\ e^{-\frac{(r-r_{\rm c})^2}{2\sigma^2}}\frac{e^{-\frac{z^2}{2(hr)^2}}}{\sqrt{2\pi}hr},
\end{equation}
with symbols having the same meaning as in Eq.~1 of \citep{matra-et-al-2018}.
The vertical aspect ratio $h=H/R$ is radially constant and fixed to 0.03 for belts that are too face-on or too low S/N for it to be meaningfully constrained. Additionally, rather than fitting for the normalization factor $\Sigma_0$, we fit for the total flux density of the belt in the model images. When calculating model images, we also assume the grains to act as blackbodies and therefore to have a temperature scaling as $r^{-0.5}$. 

After producing a model image, we Fourier Transform it and sample the model visibility function at the same \textit{u-v} locations as the data using the \textsc{GALARIO} software package \citep{tazzari-et-al-2018}. This produces model visibilities that can be directly compared with the observed ones. This process is then used to fit the model to the data through a Markov Chain Monte Carlo (MCMC) implemented using the \textsc{EMCEE} v3 software package \citep{foreman-mackey-et-al-2013, foreman-mackey-et-al-2019}. This samples the posterior probability function of the n-dimensional parameter space of our model using the affine-invariant sampler of \cite{goodman-weare-2010}. We use a likelihood function $\propto e^{-\chi^2}$ and uniform priors on all model parameters. In addition to the model parameters entering the equation describing the density distribution, we fit for RA and Dec offsets of the belt's geometric center from the phase center of the observations, and for a weight rescaling factor to account for the inaccurate data weights (and hence uncertainties) typically delivered by ALMA \citep[e.g.][]{marino-et-al-2018, matra-et-al-2019b}. We fit these additional, nuisance parameters separately to each observing date for any given target.

Tab.~\ref{tab:resolveddisc} shows best-fit parameters and uncertainties derived for each of the resolved belts, taken as the 50$^{+34}_{-34}$th percentiles of the marginalised posterior probability distribution of each parameter. Fig.~\ref{fig:ALMA_Models} shows full-resolution model images and residuals obtained by subtracting the best fit visibility model from the data, and imaging without CLEAN deconvolution. The residual images look mostly consistent with noise, indicating that our models provide a very good fit to the data.
However, we note that some residual, extended emission is detected 1) interior to the HD~181327 ring, to the SE of the star, and 2) at the SW ansa, and along the S side of the HD~191089 belt. 
While this could be due to true substructure in the dust morphology of these systems, this does not significantly affect the measurement of the radius of the bulk of the planetesimal belt material, which we are most interested in. 

\begin{table*}
\caption{Resolved discs and fitting parameters for ALMA models. 
\label{tab:resolveddisc}}
\tabcolsep 3pt
\begin{tabular}{rcccccccccc}
HD & $\lambda$ & $F_{\nu_{\star}}$ & $F_{\nu_{\rm belt}}$ & $R$  & $\Delta R$ & $h$ & $i$          & PA           & $\Delta$RA   & $\Delta$Dec \\
   & [$\mu$m]  & [$\mu$Jy]         & [mJy]                & [au] & [au]       &     & [$^{\circ}$] & [$^{\circ}$] & [$\arcsec$]  & [$\arcsec$] \\
\midrule
15115 & $1340$   & $^{a}43^{+20}_{-20}$ & $2.02^{+0.06}_{-0.06}$ & $93.4^{+1.0}_{-1.3}$ & $^{a}21^{+6}_{-7}$ & $^{a}0.051^{+0.012}_{-0.016}$ & $^{b}87.8^{+1.4}_{-1.3}$ & $98.5^{+0.3}_{-0.3}$ & $0.08^{+0.03}_{-0.03}$ & $-0.04^{+0.01}_{-0.01}$ \\
164249
 & $1350$ & $-$ & $0.96^{+0.14}_{-0.13}$ & $63^{+4}_{-3}$ & $^{a}24^{+11}_{-11}$ & $-$ & $<49$ & $^{c}113$ & $-0.08^{+0.08}_{-0.09}$ & $-0.17^{+0.08}_{-0.08}$ \\
$^\star$181327
 & $880$ & $^{a}39^{+25}_{-21}$ & $18.8^{+0.3}_{-0.3}$ & $81.3^{+0.3}_{-0.3}$ & $16.0^{+0.5}_{-0.6}$ & $<0.09$ & $30.0^{+0.5}_{-0.5}$ & $97.8^{+1.0}_{-1.0}$ & $-0.005^{+0.005}_{-0.005}$ & $-0.028^{+0.004}_{-0.004}$ \\
$^\star$191089
 & $1270$ & $^{a}45^{+21}_{-21}$ & $1.83^{+0.03}_{-0.03}$ & $44.8^{+0.9}_{-0.9}$ & $16^{+3}_{-3}$ & $^{a}0.10^{+0.04}_{-0.05}$ & $60^{+1}_{-1}$ & $73^{+1}_{-1}$ & $^{d}0.032^{+0.012}_{-0.012}$ & $^{d}-0.012^{+0.008}_{-0.008}$ \\

\bottomrule
\end{tabular}

\noindent
{\em Notes:}
$^{\star}$The model fit leaves significant residuals. 
$^a$ Marginally resolved/detected, i.e. having a posterior probability distribution with a non-zero peak but consistent with zero at the $3\sigma$ level.
$^b$ Inclination consistent with 90$^{\circ}$ (perfectly edge-on) at the 3$\sigma$ level.
$^c$ Quantity unconstrained at the 3$\sigma$ level, but with a pronounced peak at the median value reported.
$^d$ Offsets refer to 2018 dataset. For 2014 dataset, offsets were $\Delta$RA=$0.12^{+0.04}_{-0.04}$ and $\Delta$Dec=$0.02^{+0.02}_{-0.02}$.
\end{table*}

\subsection{ALMA Results}

\subsubsection{A newly resolved disc around HD~164249}
The disc around HD~164249 was observed with ALMA at 1.35~mm and is spatially resolved for the first time increasing the number of resolved debris discs reported in the literature to 153 according to the database for resolved discs\footnote{https://www.astro.uni-jena.de/index.php/theory/catalog-of-resolved-disks.html}.
It shows a face-on orientation with an inclination below 49$^\circ$. The planetesimal belt is found at 63~au with a disc width of 24~au using a Gaussian disc model. 
The disc was not resolved at any other wavelength before. 

\subsubsection{Previously resolved discs}

We re-analysed the data sets of two targets (HD~15115, HD~191089) presented in former studies to infer the system parameters, such as the disc radius, in a consistent way and present the results of new high-resolution data for HD~181327. 

HD~15115: We find the edge-on disc of HD~15115 with an inclination of 88$^\circ$ to be located at $93.4^{+1.0}_{-1.3}$~au with a disc width of $21^{+6}_{-7}$~au using a Gaussian ring model.
The results from \cite{macgregor-et-al-2015, macgregor-et-al-2019} which are based on the same dataset as our study
suggest the disc to extend from 44 to 92~au with a 14~au wide gap at 59~au.
\cite{macgregor-et-al-2019} suggests that a planet with a mass of $0.16~M_\text{Jup}$ is creating this gap, but so far no planet could be detected (see \S~\ref{sec:planets}).
Our results do not show evidence for a gap in the disc, which may be because of the different parameterisations of the two models;
\cite{macgregor-et-al-2019} assumes a 2D disc model using a power law for the radial surface density distribution and an infinitesimally small vertical scale height, whereas our disc model assumes Gaussian radial and vertical density distributions (the latter was found to be marginally resolved in our analysis).

\vspace*{0.5cm}
HD~181327: The face-on disc around HD~181327 was inferred to have a radius of $81.3\pm0.3$~au and width of $16^{+0.5}_{-0.6}$~au using a Gaussian ring model. This is comparable with the 86~au radius and width of 23~au found by \cite{marino-et-al-2016} from lower resolution ALMA Band 6 data. 

\vspace*{0.5cm}
HD~191089. The debris disc around HD~191089 was observed at 1.27~mm and formerly presented in \cite{kral-et-al-2020} which reported a disc ring at $43.4\pm2.9$\,au with a width of $<22.5$\,au and an inclination of $\sim52^\circ$.
With our Gaussian ring model we inferred an inclination of 60$^\circ$ and a disc radius of $44.8\pm0.9$~au with a width of $16\pm3$~au. The scale height was constrained to be smaller than 0.1 at a $3\sigma$ significance.
We note that our data-set does not only contain the data used in \cite{kral-et-al-2020}, but a combination of those with data from the ``Resolved ALMA and SMA Observations of Nearby Stars'' (REASONS) programme \citep[][]{sepulveda-et-al-2019} which have a higher spatial resolution, as well as older observations from 2012 (see Tab.~\ref{tab:obslog}).

\subsubsection{Gas emission}
\label{sec:gas}

We visually checked the data cubes of the four ALMA resolved targets for CO gas emission, but did not detect any. HD~181327 is the only target in our sample of F~stars in the BMPG with a gas detection presented in \cite{marino-et-al-2016}.
That study found a significant amount of $^{12}$CO in its disc based on the J=2-1 excitation level and inferred a total CO-gas mass of $1.2\ldots 2.9\times 10^{-6}M_\oplus$. The gas is consistent with a secondary origin if the planetesimals in the disc around HD~181327 possess a similar volatile fraction compared to Solar system comets. 
Our observations included the J=3-2 excitation level. The non-detection could be consistent with the J=2-1 detection depending on excitation conditions, but a full gas analysis, including optimising detection, and considering the wide range of possible excitation conditions is needed to draw a definitive conclusion.

\subsection{Imaging at other wavelengths}

\subsubsection{Scattered light and MIR observations}

Scattered light observations give us an additional opportunity to estimate the planetesimal belt radii of discs especially if they were not observed in thermal emission.
Furthermore, observations at wavelengths shorter than sub-mm trace dust grain sizes smaller than those seen with ALMA and thus can help to investigate transport processes within the discs.
While most of the spatially resolved discs in the BPMG were observed with ALMA, there is one disc (HD~160305) only observed in scatterd light.

HD~160305: 
The disc around HD~160305 was recently detected with VLT/SPHERE by \cite{perrot-et-al-2019} in scattered light. The debris dust is confined to a narrow ring between 86 and 90~au. It shows a near edge-on inclination and a brightness asymmetry between its western and eastern side. \cite{perrot-et-al-2019} suggest different scenarios as the reason for this asymmetry, such as strong recent collisions of planetesimals, interactions with massive companions, or pericentre glow effects, but was not able to differentiate between these scenarios.

HD~15115: Scattered light observations of HD~15115 \citep[e.g.,][]{kalas-et-al-2007,engler-et-al-2019} revealed a strong asymmetry of the disc which is not seen in ALMA observations. 
\cite{kalas-et-al-2007} report a disc extent up to 580~au on the west side and 340~au on the east side. \cite{macgregor-et-al-2019} concluded that the mechanism causing the asymmetry is only affecting the smallest grains, suggesting interaction with the local ISM as a likely reason for it. \cite{engler-et-al-2019} derived the maximum of polarised flux density at a location of $94\pm2$~au, which is assumed to correspond to the location of the planetesimal belt (similar to the radius we find in Tab.~\ref{tab:resolveddisc}). 

HD~181327: \textit{HST}/NICMOS observations of HD~181327 in scattered light presented by \cite{schneider-et-al-2006} derived a disc radius of 86~au with a width of 36~au. While the radius is in agreement with our ALMA-based results the disc width is broader in scattered light than at sub-mm wavelengths. We would expect a broader disc at shorter wavelengths since such observations trace smaller particles which are more susceptible to transport processes.  
Asymmetries were reported by \cite{stark-et-al-2014} which suggested a recent catastrophic disruption or a warping of the disc by the ISM as probable causes.

HD~191089: 
\cite{churcher-et-al-2011} observed HD~191089 at 18.3$\mu$m with T-ReCS on Gemini South and found excess emission between 28 and 90~au. This is in agreement with the belt location inferred from observations of the HD~191089 disc performed by \textit{HST}/NICMOS and STIS and \textit{Gemini}/GPI \citep{ren-et-al-2019}. That study detected scattered light between 26 and 78~au. In addition to the dust ring a halo was found extending to $\sim640$~au, but no brightness asymmetries were identified.
However, similar to HD~181327 the disc is broader in scattered light than at mm wavelengths.

\subsubsection{FIR observations with \textit{Herschel}/PACS}
\label{sec:herschel}

Three discs in the BPMG sample were spatially resolved in the FIR with \textit{Herschel}/PACS (HD~15115, HD~164249, and HD~181327). 
To infer their radii in a homogeneous way we apply the method described in \cite{yelverton-et-al-2019} and \cite{xuan-et-al-2020}. The PSF of the star is derived from observations of the \textit{Herschel} calibration star HD~164058 and then rotated to the appropriate orientation and scaled to the stellar flux derived from the SED.
Then we generate an axisymmetric, optically thin disc model and assume that the surface brightness is proportional to $r^{-1.5}$ with $r$ being the distance to the star. The dust temperature at a distance $r$ is assumed to follow $r^{-0.5}$ as for blackbody grains. The free parameters of the model are the total disc flux density, the inner and outer edges of the disc, $R_\text{in}$ and $R_\text{out}$, inclination, and position angle. We also include a 2D offset to account for non-perfect \textit{Herschel} pointing. The model disc is convolved with the PSF and then compared to the \textit{Herschel}/PACS image by calculating the goodness of fit, $\chi^2$.
We estimate the parameters using the emcee package.
To get disc radii to be compared with those inferred from ALMA images that assumed a Gaussian radial profile, we derive a radius from \textit{Herschel}/PACS, $R_\text{FIR} = 0.5\times (R_\text{in} + R_\text{out})$. We note that in some cases the inner and outer edge of a disc might be poorly constrained, and in any case $R_\text{FIR}$ could differ from the location of the peak of the surface brightness that would have been inferred if observed at higher spatial resolution. 

The modelling results of the \textit{Herschel}/PACS images for the BPMG sample are listed in Tab.~\ref{tab:herschel}. In all cases the inclination and position angle are in agreement with ALMA observations listed in Tab.~\ref{tab:resolveddisc}.  For HD~15115 and HD~181327 the radii,  $R_\text{FIR}$, are in agreement with values inferred from ALMA observations, but possess uncertainties of up to 25\%. The disc widths seem to be larger compared to ALMA, but only show deviations within $2\sigma$ so that we assume the discs to be similar in ALMA and \textit{Herschel}. A broader extent in \textit{Herschel} might indicate the presence of transport mechanisms altering the orbits of smaller dust particles towards larger eccentricities. 
For HD~164249 the \textit{Herschel} results show large uncertainties due to the low spatial resolution. The \textit{Herschel} radius is a factor 2.2 smaller than that from ALMA, however the two radii are not significantly different ($\sim2\sigma$) and the higher resolution ALMA result is a better estimate of the planetesimal belt location.

\begin{table*}
    \caption{Discs resolved with \textit{Herschel}/PACS.
    \label{tab:herschel}}
    \centering
    \tabcolsep 3pt
    \begin{tabular}{rcccccc}
    HD & $R_\text{in}$    & $R_\text{out}$ & $R_\text{FIR}$ & $\Delta R_\text{FIR}$ & $i$          & PA \\
       & [au]   & [au] & [au] & [au] & [$^\circ$]   & [$^\circ$] \\
    \midrule
     15115 & $40.6\pm23.7$ & $145.4\pm29.0$ & $93.0\pm18.7$ & $52.4\pm18.7$ & $85.9\pm3.7$ & $98.9\pm2.4$\\
    164249 & $13.3\pm11.5$ & $43.0\pm21.2$ & $28.2\pm12.0$ &$14.8\pm12.0$& $68.1\pm20.8$ & $175.1\pm66.5$\\
    181327 & $25.7\pm25.5$ & $134.4\pm30.8$ & $80.0\pm20.0$ &$54.3\pm20.0$& $30.1\pm8.9$ & $101.6\pm11.6$\\
    \bottomrule
\end{tabular}

\noindent
{\em Notes:}
    $R_\text{in}$ and $R_\text{out}$ give the inner and outer radii for \textit{Herschel}/PACS inferred with the method described in \S~\ref{sec:herschel} following the procedure of \cite{yelverton-et-al-2019} and \cite{xuan-et-al-2020}.
    $R_\text{FIR}$ is the central radius defined as $R_\text{FIR}=0.5*(R_\text{in} + R_\text{out})$, $\Delta R_\text{FIR}$ gives the disc width. The parameters $i$ and PA give the inclination and position angle.
\end{table*}

\subsection{Ratio of spatially resolved disc radius to blackbody radius}
\label{sec:trueradius}

\begin{figure}
\centering
\includegraphics[width=\columnwidth]{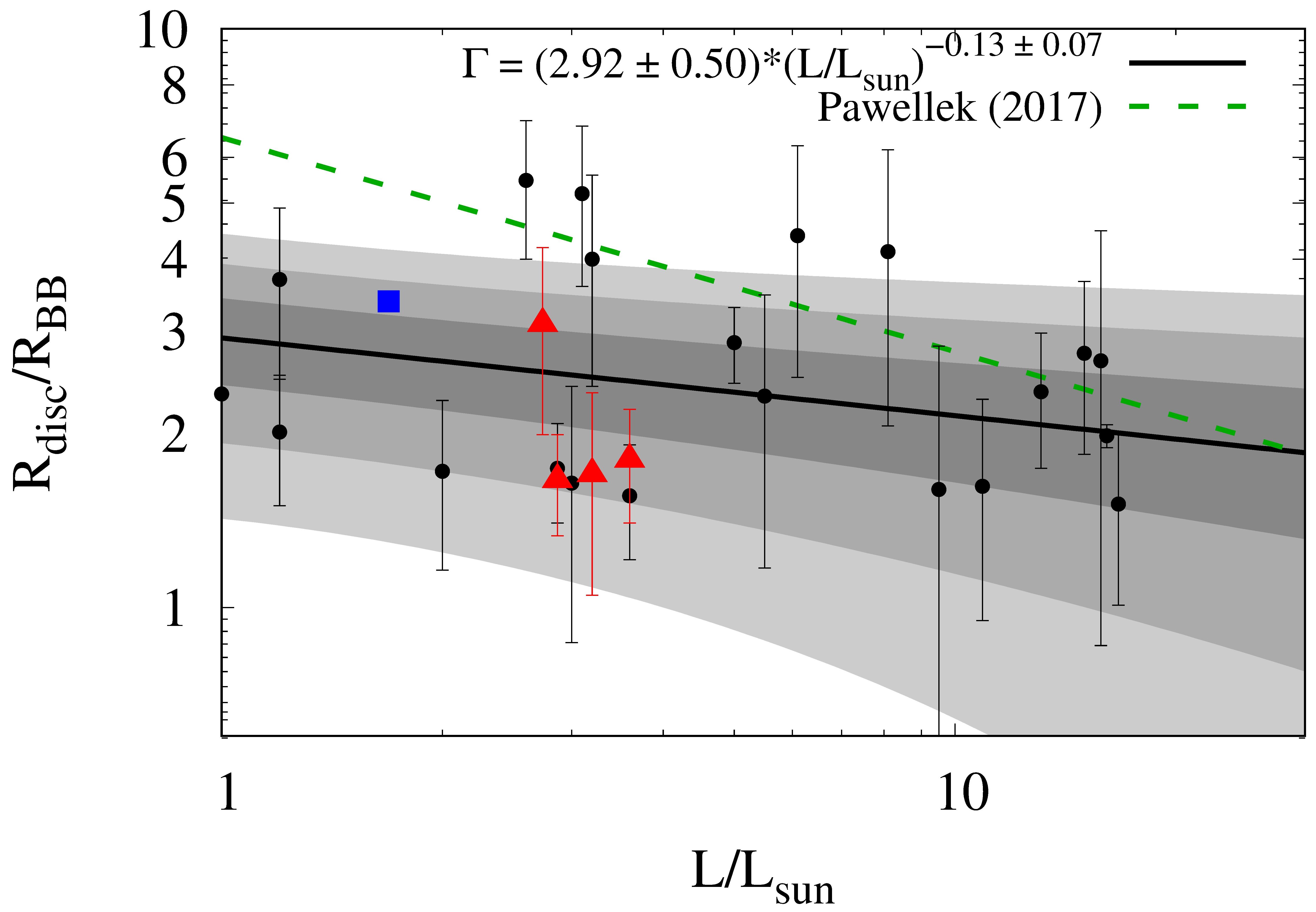}
 \caption[]{\label{fig:Rv_L} Spatially resolved disc radius to blackbody radius ratio as a function of stellar luminosity for NIR, FIR, and sub-mm wavlengths. Black circles show the ALMA sample from \protect{\cite{matra-et-al-2018}}, red asterisks show the ALMA-resolved F~stars and the blue square the SPHERE-resolved F~star HD~160305. The grey shaded areas depict the 1, 2, and $3\sigma$ levels of the correlation. The green dashed line shows the trend found in \protect{\cite{pawellek-2017}} from the Herschel resolved disc radius.}
\end{figure}

We calculate the ratio of sub-mm radius to blackbody radius for the four ALMA resolved discs in our sample. In addition, we infer the ratio of the scattered light radius to blackbody radius for HD~160305. Given the knowledge that there is a potential trend in sub-mm disc sizes with stellar luminosity \citep[][]{matra-et-al-2018}, and also a trend in the far-infrared size to blackbody radius ratio with stellar luminosity \citep{booth-et-al-2013, pawellek-krivov-2015}, we compare the five discs to a sample of ALMA-resolved discs with a broader stellar luminosity range \citep{matra-et-al-2018}. 
In Fig.~\ref{fig:Rv_L} we plot the radius ratio as a function of stellar luminosity. The actual disc width inferred by ALMA observations (see Tab.~\ref{tab:resolveddisc}) is given by the error bars.

For our sample of F~stars the values of this ratio lie between 1.6 and 3.4 where the system with the lowest stellar luminosity (HD~160305) possesses the highest value.
Including the ALMA-sample of \cite{matra-et-al-2018} 
and fitting a trend of the form in eq.~\ref{eq:trueradius} for how the ratio depends on stellar luminosity we infer a slight decrease of the ratio with stellar luminosity 
finding parameter values of $A = 2.92\pm0.50$ and $B=-0.13\pm0.07$. 

This result is different from the trend presented in \cite{pawellek-krivov-2015} and \cite{pawellek-2017} based on disc sizes from \textit{Herschel} images which showed parameter values of $A = 6.49\pm0.86$ and $B=-0.37\pm0.05$ (see green dashed line in Fig.~\ref{fig:Rv_L}). 
For systems with stellar luminosities larger than $5L_\text{sun}$ the radius ratio of the ALMA sample is in agreement with \cite{pawellek-krivov-2015}. The different fit is caused by a number of systems with lower luminosities including our sample of ALMA resolved F~stars that show relatively small ratios.
Possible reasons for the different trends could be that \textit{Herschel} had a lower resolution and so there may be systematic uncertainties in the derived disc radii, or the discs could be larger when traced in the far-IR due to such wavelengths tracing small dust in the halo that extends beyond the planetesimal belt.
However, our analysis of the \textit{Herschel} images of the BPMG F~stars in \S~\ref{sec:herschel} inferred radii that are consistent with those from ALMA images. Considering \cite{pawellek-krivov-2015}, none of the BPMG targets was used to derive the radius ratio vs luminosity trend, but the study inferred radii between 93 and 112~au for HD~181327 depending on the dust composition assumed which is in agreement with the results of ALMA and \textit{Herschel}.
A more detailed analysis is needed to investigate possible causes for the different outcomes between the \textit{Herschel} and ALMA samples. 
A systematic difference might indicate the presence of dynamical processes affecting the size distribution in a way not considered before.

\section{SED modelling revisited}
\label{sec:SEDmodelling}

As mentioned before, five discs of our sample were spatially resolved (four with ALMA and one with SPHERE, see Tab.~\ref{tab:resolveddisc}). 
This allows us to apply a more detailed model to fit the SEDs of these five discs rather than using a simple modified blackbody model as in \S\ref{sec:IRExcess}. 
In the following approach we model the dust size distribution and composition.

\subsection{Modelling approach}

We  use  the SONATA code  \citep[][]{pawellek-et-al-2014, pawellek-krivov-2015} and apply the same PHOENIX-GAIA stellar photospheric models \citep[][]{brott-hauschildt-2005} to determine the host star contribution in a similar approach as for the modified blackbody fits (MBB, see \S\ref{sec:IRExcess}). 
While for the MBB model we simply fitted a dust temperature and a fractional luminosity without consideration of dust properties the SONATA code calculates the temperature and the thermal emission of dust particles at different distances to the star. It assumes compact spherical grains and uses Mie theory to derive the absorption efficiencies \citep{bohren-huffman-1983}. The dust composition is assumed to be pure astronomical silicate \citep[][]{draine-2003} with a bulk  density of 3.3~g/cm$^{3}$.
The code sums up the emission of particles within a range of sizes to generate the SED. 
The flux densities given for wavelengths shorter than 5~$\mu$m are not used to fit the dust disc since in this wavelength regime the stellar photosphere rather than the dust dominates the emission. 

We apply a power law for the size distribution and assume a Gaussian radial distribution of the dust using the surface number density $N(r,s)$:
\begin{equation}
 N_\text{SED}(r,s)\sim s^{-q_\text{SED}} \frac{1}{\sqrt{2\pi}\Delta R_\text{disc}}\exp\left[-\frac{1}{2}\left(\frac{r-R_\text{disc}}{\Delta R_\text{disc}}\right)^2\right].
\end{equation}
Here, $r$ represents the distance to the star, $R_\text{disc}$ the peak and $\Delta R_\text{disc}$ the width of the radial distribution. The parameter $s$ is the grain radius and $q_\text{SED}$ is the SED power-law index for the size distribution. The surface number density is directly connected to the surface density, $\Sigma$, by $\Sigma(r,s)\,ds=\pi s^2\, N(r,s)\,ds$.

The grain sizes lie between a minimum and a maximum value, $s_\text{min}$ and $s_\text{max}$ where we fix the maximum grain size to 10~cm. Larger grains do not contribute to the SED in the wavelength range observed for the size distributions considered here with $q_\text{SED}>3$. In addition, we fix the radial parameters to the values inferred from our resolved images (see Tab.~\ref{tab:resolveddisc}). 
Therefore, we are left with three free parameters to fit: the minimum grain size, $s_\text{min}$, the size distribution index, $q_\text{SED}$, and the amount of dust, $M_\text{dust}$, for particles between $s_\text{min}$ and $s_\text{max}$ assuming a bulk density $\varrho$.

We follow the three criteria given in \cite{ballering-et-al-2013} and \cite{pawellek-et-al-2014} to check for the presence of a warm component for the five discs.
For us to consider a warm component to be present, there has to be a significant excess ($\ge3\sigma$) in either the WISE/22 or MIPS/24 
in excess of that which could originate in a single ring fitted to longer wavelength data.
Secondly, the fit of the two-component SED has to be much better than the one-component fit. While the former studies assumed a better two-component fit when $\chi^2_\text{one}/\chi^2_\text{two}>3$ was fulfilled we use the Bayesian information criterion (BIC) instead, which is  
\begin{equation}
\text{BIC} = \chi^2 + J \log_e{(N)},
\end{equation}
where $J$ represents the number of free parameters and $N$ the number of data points. We use the classification given in \cite{kass-raftery-1995} to infer whether a one- or a two-component model is more likely.
As a third criterion we require the inferred ring containing the warm dust to be located outside the sublimation radius $R_\text{sub}$ (assuming 1300~K as the sublimation temperature).

If all three criteria are fulfilled we obtain the two-component model in the following way.
In a first step we assume the warm dust to be modelled by a pure blackbody to infer its blackbody temperature and radius.
We assume this radius to be the location of the warm dust belt and fix the belt width to $\Delta R_\text{disc}/R_\text{disc}=10\%$. Finally, we fit both disc components assuming that
the warm and cold dust ring possess the same size distribution of dust grains. 

Similar to the cold dust ring it is likely that the sub-mm disc radius of the warm belt is larger than the blackbody radius. 
Applying the newly inferred values presented in \S~\ref{sec:trueradius} the factor would be $\sim2.5$, but it could be smaller or larger, since a consistently different dust temperature or composition could result in a systematic difference.  

For the disc around HD~160305 only four mid- and far-infrared data points (WISE12, WISE22, PACS70, PACS160) are listed in the literature. Therefore, we fix the size distribution index, $q_\text{SED}$, to 3.5 \citep[the outcome of an ideal collisional cascade, ][]{dohnanyi-1969} to reduce the number of free parameters.

\subsection{Fitting results}

Following the criteria for two-component models we checked at first the SEDs of the four resolved discs around HD~15115, HD~164249, HD~181327 and HD~191089 for the presence of a warm inner component. Only HD~15115 fulfills all of them so that we fitted this disc with a two-component model.
The SED fitting results of the whole sample are all summarised in Tab.~\ref{tab:SEDresults} and the SEDs are depicted in Fig.~\ref{fig:SEDs}.

Collisional evolution models show that grains smaller than a certain blow-out size, $s_\text{blow}$, are expelled from the stellar system due to radiation pressure. The blow-out size depends on the optical parameters of the dust material and increases with increasing stellar luminosity. 
We would expect the minimum grain size, $s_\text{min}$, to be comparable to $s_\text{blow}$. However, previous studies of grain size distributions \citep[e.g.,][]{pawellek-et-al-2014, pawellek-krivov-2015} found that $s_\text{min}$, is weakly connected to the stellar luminosity. It might also be consistent with being independent of stellar luminosity, since those studies found an average value of $\sim5\mu$m to fit the majority of debris discs analysed therein. 
It was also found that the ratio between $s_\text{min}$ and $s_\text{blow}$ is $\sim 4 \ldots 5$ for discs around host stars with stellar luminosities between 2 and 5 $L_\text{sun}$ \citep[][]{pawellek-et-al-2014}.  
The $s_\text{min}/s_\text{blow}$ ratio is thought to be connected to the dynamical excitation of the planetesimals producing the visible dust \citep[e.g.,][]{krijt-kama-2014, thebault-2016}. Earlier studies, such as \cite{krivov-et-al-2006} or \cite{thebault-augereau-2007} suggest a value around 2 for collisionally active discs. 

For three targets in our sample our modelling finds that $s_\text{min}$ is close to $s_\text{blow}$ leading to a $s_\text{min}/s_\text{blow}$ ratio of $\sim1$.
Only the results for HD~15115 reveal a $s_\text{min}$ close to $5\mu$m and a $s_\text{min}/s_\text{blow}$ ratio of $\sim5$. However, the difference in $s_\text{min}$ of this disc to the others in the sample should be treated with caution, since the minimum grain size that we infer may be influenced by how we treated the warm component that is only present in this system. Besides our four targets there is a range of different discs at the same stellar luminosity investigated by \cite{pawellek-krivov-2015} and \cite{matra-et-al-2018} and shown as black dots in Fig.~\ref{fig:Rv_L}, most of which possess a larger $s_\text{min}$. 
The low $s_\text{min}/s_\text{blow}$ ratio for F~stars in the BPMG, which is reported for the first time, could indicate high levels of dynamical excitation similar to that found for discs around A-type stars \citep[see Fig.~16 in ][]{pawellek-krivov-2015}. 

The size distribution index, $q_\text{SED}$, lies between 3.4 and 3.8 for our sample.
These values are consistent with collisional models \citep[e.g.,][]{loehne-et-al-2007, gaspar-et-al-2012, kral-et-al-2013, loehne-et-al-2017}.

Overall, the results from our SED modelling suggest that all four spatially resolved discs are in agreement with a stirred debris disc scenario which means that
the dust seen in the SED is consistent with being created by the collisional destruction of planetesimals in a belt traced by the ALMA images.

\section{Comparison with nearby F stars}
\label{sec:comparison}

In the first part of this study we analysed the properties of the BPMG in detail.
So far we do not know whether the high incidence rate of debris discs is a peculiarity of said moving group or whether we see more discs due to the young age of the moving group. Therefore, we will put the results of the BPMG into context with discs around other near-by F-type stars in the second part of this study.
First we investigate the evolution of spectral type to ensure that we compare stellar populations with similar properties. Then we look at the appropriate systems in samples of field stars and other young moving groups.

\subsection{Stellar population at different ages}

The stellar spectral type is determined by the effective temperature of the star.
Due to ongoing thermonuclear reactions, stars and their physical/chemical properties such as metallicity, stellar radius or temperature, evolve over time so that the spectral type might change as well. 
Therefore, it is not self-evident that comparing stars with similar spectral types but different ages show the same stellar population at varying evolutionary phases. 
 
\begin{figure}
 \includegraphics[angle = -90, width=\columnwidth]{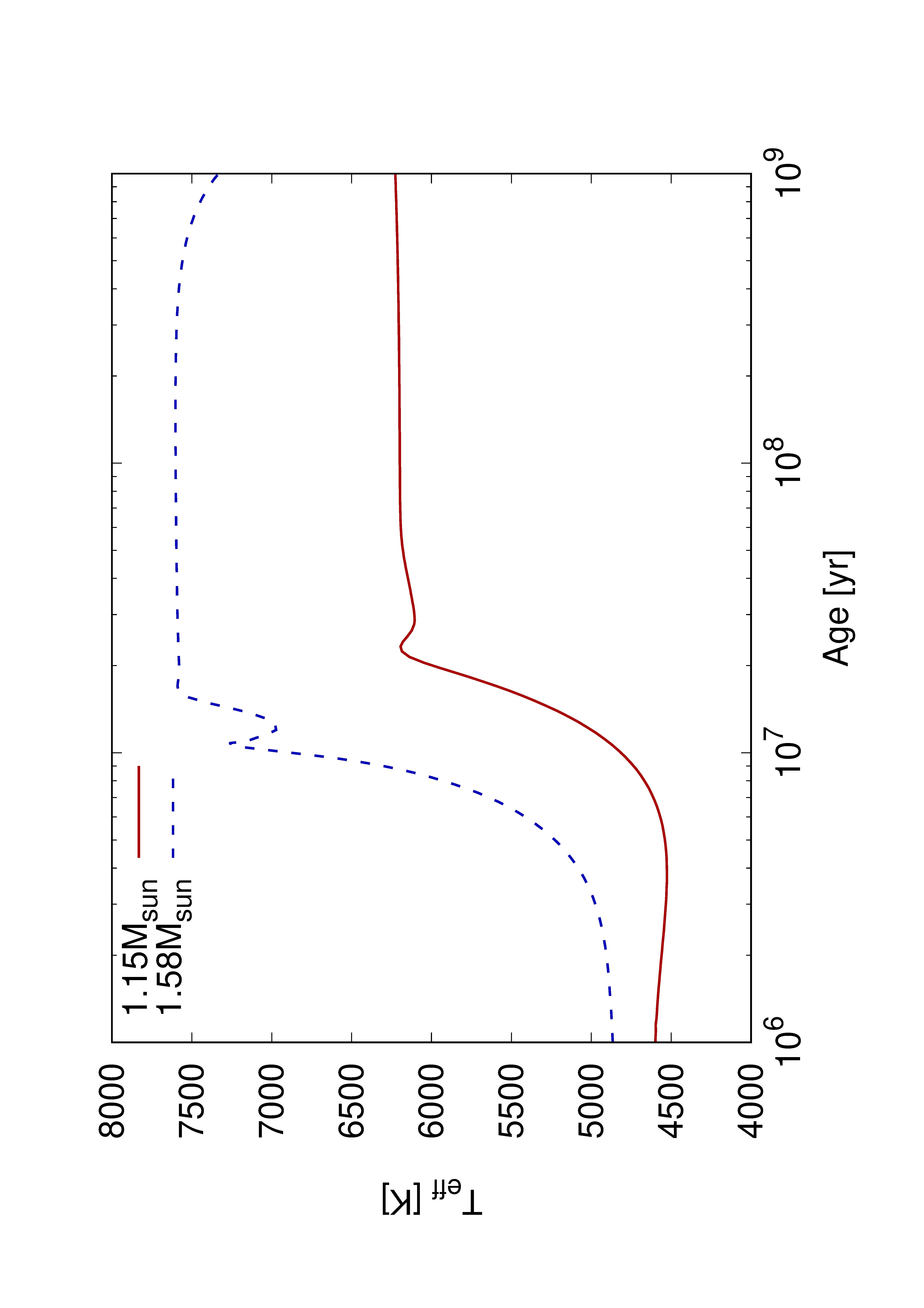}
 \caption{Stellar temperature as function of age.
 \label{fig:EvolutionModel}}
\end{figure}

We use the "`Modules for Experiments in Stellar Astrophysics"' \citep[MESA, ][]{paxton-et-al-2011, paxton-et-al-2013, paxton-et-al-2015, choi-et-al-2016} to check the evolution of stellar temperature over time. MESA consists of a one-dimensional stellar evolution module simultaneously solving the fully coupled structure and composition equations.
The results are shown in Fig.~\ref{fig:EvolutionModel}. 
We use the lowest ($1.15M_\text{sun}$) and highest ($1.58M_\text{sun}$) stellar masses in our sample of F stars to analyse its parameter space and assume a stellar metallicity of $[\text{Fe}/\text{H}]=0.0$. 

The stellar temperature increases up to an age of $\sim10$~Myr and then stays constant until $\sim1$~Gyr. Our sample of F stars belongs to the BPMG with an age of 23~Myr. Fig.~\ref{fig:EvolutionModel} shows that at this age the temperature is already constant so that the spectral type is not changing. As a result, stars with similar spectral types and ages between that of the BPMG and 1~Gyr should represent the same population of stars. 
As the stars leave the main-sequence from their position in the HR diagram the stellar temperature starts to decrease. Higher mass stars (e.g., that were A~stars on the main sequence) evolve to become F~stars as they leave the main sequence. 
Therefore, a sample of F~stars may be contaminated by post-main sequence (higher-mass) F~stars. Their fraction should be small in a volume-limited population since the number of high mass stars is lower. Furthermore, those stars do not spend long on the post-main sequence looking like an F~star before they noticeably evolve so that they should be possible to identify from their stellar luminosity.

\subsection{Field stars}
\label{sec:fieldstars}

\cite{sibthorpe-et-al-2018} analysed an unbiased sample of 275 FGK~stars including 92 F-type stars observed with \textit{Herschel}/PACS in the DEBRIS survey. None of the F-types belong to the BPMG, which lie within 24~pc.
All targets are older than 160~Myr following the age determination of \cite{vican-2012} with the exception of the targets HD~56986 with an age of $\sim20$~Myr and HD~7788A where no age is given.

Based on \cite{sibthorpe-et-al-2018}, 22 of the 92 stars show evidence for a debris disc. However, we note that the target HD~19994 \citep{wiegert-et-al-2016} previously assumed to have spatially resolved IR emission shows evidence of being confused rather than possessing an actual disc \citep{yelverton-et-al-2019}. Thus, we update the number of detections for F~stars in the DEBRIS sample to 21 out of 92 targets leading to a detection rate of $22.8^{+6.2}_{-4.9}$\%. Due to the large beam size of \textit{Herschel}/PACS other targets might show confusion as well. However, after checking the PACS images available we did not identify more potentially confused discs.
The HR-diagram of the whole DEBRIS sample is presented in Figs.~1 and 2 in \cite{phillips-et-al-2010} and shows that all F~stars in the DEBRIS sample which possess debris discs are compatible with belonging to the main sequence. Therefore, we assume that in the DEBRIS sample no debris discs around post-main sequence F~stars are included.

\begin{table}
\caption{F~stars in the DEBRIS sample with debris disc detections.
\label{tab:DEBRIS_sample}}
\tabcolsep 2pt
\scriptsize
\centering
\begin{tabular}{rccrclcrclcc}
HD & d & SpT & \multicolumn{3}{c}{$L/L_\text{sun}$} & $f_\text{d}$ & \multicolumn{3}{c}{$R_\text{BB}$} & $R_\text{\text{FIR}}$ & $\Delta R_\text{\text{FIR}}$ \\
   & [pc] &  & \multicolumn{3}{c}{ } & [$10^{-5}$] & \multicolumn{3}{c}{[au]} & [au] & [au]\\
\midrule
  1581 &  8.6 & F9.5V      & 1.29 & $\pm$ & 0.01 & 0.05 & 124 & $\pm$ & 45 & \ldots & \ldots\\
  7570 & 15.2 & F9VFe+0.4  & 1.96 & $\pm$ & 0.01 & 1.20 &  28 & $\pm$ & 10 & \ldots &\ldots\\
 $^*$10647 & 17.3 & F9V        & 1.55 & $\pm$ & 0.01 & 32.7 &  25 & $\pm$ &  6 & $112.3\pm+2.5$     & $69.4\pm2.5$ \\
 11171 & 23.2 & F0V        & 5.80 & $\pm$ & 0.10 & 0.68 &  32 & $\pm$ & 17 & \ldots &\ldots\\
 16673 & 21.9 & F8VFe-0.4  & 1.93 & $\pm$ & 0.02 & 0.33 &  33 & $\pm$ & 23 & \ldots & \ldots\\
 $^*$22484 & 14.0 & F9IV-V     & 3.22 & $\pm$ & 0.06 & 0.72 &  21 & $\pm$ &  8 & $39.7\pm20.5$ & $29.4\pm20.5$\\
 $^*$27290 & 20.5 & F1V        & 6.67 & $\pm$ & 0.09 & 1.53 &  77 & $\pm$ & 23 & $151.2\pm32.2$ & $121.1\pm32.2$\\
 33262 & 11.6 & F9VFe-0.5  & 1.47 & $\pm$ & 0.01 & 1.29 &  6.1& $\pm$ & 2.9& \ldots & \ldots\\
 $^*$48682 & 16.7 & F9V        & 1.86 & $\pm$ & 0.02 & 4.74 &  69 & $\pm$ & 16 & $134.4\pm7.6$ & $73.7\pm7.6$ \\
 55892 & 21.4 & F3VFe-1.0  & 5.68 & $\pm$ & 0.07 & 1.21 & 2.1 & $\pm$ & 1.5 & \ldots & \ldots\\
 $^a$56986 & 18.5 & F2VkF0mF0  & 11.8 & $\pm$ & 0.20 & 160 & 0.1& $\pm$ & 0.1 & \ldots &\ldots\\
 $^*$90089 & 22.7 & F4VkF2mF2  & 3.31 & $\pm$ & 0.04 & 1.01 & 140 & $\pm$ & 37 & $58.1\pm30.8$ & $34.9\pm30.8$\\
102870 & 10.9 & F9V        & 3.73 & $\pm$ & 0.04 & 0.06 &  48 & $\pm$ & 68 & \ldots &\ldots\\
$^*$109085 & 18.3 & F2V        & 4.85 & $\pm$ & 0.09 & 2.76 &  67 & $\pm$ & 18 & $150.4\pm10.6$ & $56.7\pm10.6$ \\
$^*$110897 & 17.6 & F9VFe-0.3  & 1.11 & $\pm$ & 0.01 & 1.89 &  52 & $\pm$ & 15 & $97.14\pm48.3$ & $65.7\pm48.3$ \\
128167 & 15.7 & F4VkF2mF1  & 3.22 & $\pm$ & 0.03 & 1.38 &  8.3& $\pm$ & 22 & \ldots &\ldots\\
160032 & 21.2 & F4V        & 4.55 & $\pm$ & 0.06 & 0.29 &  64 & $\pm$ & 35 & \ldots &\ldots\\
$^*$165908 & 15.6 & F7VgF7mF5  & 2.87 & $\pm$ & 0.07 & 0.80 & 150 & $\pm$ & 36 & $138.5\pm40.8$ & $64.4\pm40.8$ \\
199260 & 21.3 & F6V        & 1.97 & $\pm$ & 0.01 & 1.57 &  26 & $\pm$ & 12 & \ldots & \ldots\\
$^*$219482 & 20.5 & F6V        & 1.90 & $\pm$ & 0.01 & 3.26 &  15 & $\pm$ &  6 & $20.6\pm12.2$ & $12.3\pm12.2$ \\
222368 & 13.7 & F7V        & 3.33 & $\pm$ & 0.03 & 0.98 &  5.9& $\pm$ & 6.7& \ldots &\ldots\\
\bottomrule
\end{tabular}

\noindent
{\em Notes:}
$^*$Target was reported in \cite{sibthorpe-et-al-2018} to possess extended disc emission. 
The radii, $R_\text{FIR}$, from \textit{Herschel}/PACS were derived from the model presented in \cite{yelverton-et-al-2019} and are defined in the same way as described in \S\ref{sec:herschel}.
$^a$HD~56986 possesses a marginal excess at MIPS24. The image at PACS160 seems to be confused with a nearby background object making the SED model very uncertain.
\end{table}

The SEDs are fitted using the same process outlined in \S~\ref{sec:herschel}. The
modelling results are listed in Tab.~\ref{tab:DEBRIS_sample} and the SEDs are shown in Figs.~\ref{fig:SEDs_DEBRIS_1} and \ref{fig:SEDs_DEBRIS_2}. We find blackbody radii between 2 and 200~au for the whole sample with the exception of HD~56986 with a blackbody radius around 0.1~au based on a marginal mid-IR excess. The excess found at PACS160 is confused by a nearby background object so that the SED model is very uncertain. We therefore exclude this target from our further analysis.

Ignoring HD~56986 due to the aforementioned reasons, we find no discs smaller than 1~au, one disc out of 92 targets with a blackbody radius between 1 and 3~au (1.1\%), three disc radii between 3 and 10~au (3.3\%), five discs between 10 and 30~au (5.4\%), seven discs between 30 and 100~au (7.6\%), and four discs larger than 100~au (4.3\%). 
Nine targets were reported to be spatially resolved in the FIR \citep{sibthorpe-et-al-2018} (excluding HD~19994). Only HD~10647 and HD~109085 were observed with ALMA (see \S~\ref{sec:planetesimalsizes}). However, using the method of \cite{yelverton-et-al-2019} we infer radii and disc widths from \textit{Herschel}/PACS images in the same manner as described in \S~\ref{sec:herschel} (see Tab.~\ref{tab:DEBRIS_sample}).
The discs range from 20~au to more than than 150~au. The smallest discs are located around HD~22484 and HD~219482 with radii of 39.7 and 20.6~au respectively. 
The disc widths are uncertain because of the relatively poor spatial resolution so that we cannot draw strong conclusions on them.

\subsection{Other young moving groups}
\label{sec:youngMG}

The question arises whether the high occurrence rate of debris discs around F-type stars in BPMG is a singular phenomenon of this moving group or if it is common in other associations with comparable properties in age and distance as BPMG. 
Here we compare the BPMG disc incidence rates with those of other clusters. When doing so we need to recognise that some stars lack FIR data and so have limited constraints on the presence of circumstellar dust.
We will consider detection rates for the whole sample (e.g. the 9/12 rate from the BPMG) and separately we will consider the rate amongst those with FIR data (e.g. the 9/11 rate for the BPMG).

Following studies of young associations \citep[e.g., Fig.~7 in][]{gagne-et-al-2018, gagne-faherty-2018} we identified five groups with similar peaks in their distance distributions around 50~pc comparable to the BPMG: the Tucana/Horologium association (THA), Columba (COL), Carina (CAR), AB~Doradus (ABDMG) and Carina-Near (CARN). 
The groups THA, COL and CAR possess an age around $\sim$45~Myr, the groups ABDMG and CARN an age around $\sim150$~Myr. 
For the purpose of our analysis a differentiation between the single moving groups is not necessary. Indeed, \cite{torres-et-al-2008} and \cite{zuckerman-et-al-2011} found that THA, COL and CAR
are closely located making it difficult to place members in one or the other group.  
Therefore, we generated two samples, one referred to as 45~Myr group sums up all F-type targets belonging to THA, COL and CAR, the other referred to as 150~Myr group combines the targets of ABDMG and CARN. Both samples are unbiased towards the presence of IR-excess. 

\begin{table}
\caption{F~stars of the 45 and 150~Myr groups.
\label{tab:youngstars}}
\tabcolsep 2pt
\centering
\scriptsize
 \begin{tabular}{rccccccccrcl}
HD     & Group & d    & SpT & \multicolumn{3}{c}{$L/L_\text{sun}$} & Disc   & $f_\text{d}$  & \multicolumn{3}{c}{$R_\text{BB}$} \\
       &       & [pc] &     & \multicolumn{3}{c}{}                 & excess & [$10^{-5}$]   & \multicolumn{3}{c}{[au]}          \\
  \midrule
   984 & 45~Myr    & 45.9	& F7V      & 2.04 &$\pm$& 0.02 & No  & -    & \multicolumn{3}{c}{-} \\
  1466 & 45~Myr    & 43.0	& F8V      & 1.58 &$\pm$& 0.01 & Yes & 6.3  & 7.8 &$\pm$& 1.8       \\
  8671 & 45~Myr    & 42.7  & F7V      & 6.08 &$\pm$& 0.04 & \ldots  & \ldots    & \multicolumn{3}{c}{\ldots} \\
 10269 & 45~Myr    & 46.7  & F5V      & 2.60 &$\pm$& 0.02 & Yes & 12   & 8.8 &$\pm$& 5.2       \\
 10863 & 45~Myr    & 45.0  & F2V      & 4.39 &$\pm$& 0.03 & No  & -    & \multicolumn{3}{c}{-} \\
 12894 & 45~Myr    & 46.3	& F4V      & 4.50 &$\pm$& 0.04 & No  & -    & \multicolumn{3}{c}{-} \\
 13246 & 45~Myr    & 45.6	& F7V      & 1.72 &$\pm$& 0.04 & Yes & 14  & 5.4 &$\pm$& 1.4         \\
 14691 & 45~Myr   & 30.0  & F3V      & 4.76 &$\pm$& 0.04 & No  & -    & \multicolumn{3}{c}{-} \\
 17250 & 45~Myr    & 57.1	& F8       & 1.91 &$\pm$& 0.02 & \ldots  & \ldots    & \multicolumn{3}{c}{\ldots} \\
 20121 & 45~Myr    & 42.5	& F3V+A8V  & 5.70 &$\pm$& 0.60 & \ldots & \ldots    & \multicolumn{3}{c}{\ldots} \\
 20385 & 45~Myr   & 48.8	& F6V      & 2.08 &$\pm$& 0.02 & No  & -    & \multicolumn{3}{c}{-} \\
 21024 & 45~Myr    & 29.3  & F5IV-V   & 4.25 &$\pm$& 0.03 & \ldots & \ldots    & \multicolumn{3}{c}{\ldots} \\
 24636 & 45~Myr    & 57.1	& F3IV/V   & 3.66 &$\pm$& 0.02 & Yes & 9.9   & 13 &$\pm$& 3      \\
 29329 & 45~Myr    & 32.7  & F7V      & 2.25 &$\pm$& 0.02 & \ldots & \ldots    & \multicolumn{3}{c}{\ldots} \\
 30051 & 45~Myr    & 67.6  & F2/3IV/V & 5.20 &$\pm$& 0.20 & Yes & 3.0  & 48   &$\pm$& 19       \\
 30132 & 45~Myr    & 121   & F6/7V    & 3.03 &$\pm$& 0.03 & \ldots & \ldots    & \multicolumn{3}{c}{\ldots} \\
 30447 & 45~Myr    & 80.5  & F3V      & 3.68 &$\pm$& 0.03 & Yes & 92  & 34  &$\pm$& 8         \\
 30984 & 45~Myr    & 82.6  & F5V      & 2.09 &$\pm$& 0.02 & \ldots  & \ldots    & \multicolumn{3}{c}{\ldots} \\
 31359 & 45~Myr    & 112   & F5V      & 3.30 &$\pm$& 0.20 & \ldots  & \ldots    & \multicolumn{3}{c}{\ldots} \\
 32195 & 45~Myr    & 62.8	& F7V      & 1.34 &$\pm$& 0.01 & Yes & 8.5  & 14   &$\pm$& 7        \\
 35114 & 45~Myr    & 47.7	& F6V      & 2.08 &$\pm$& 0.02 & Yes & 4.0  & 6.4  &$\pm$& 2.7    \\
 35996 & 45~Myr    & 92.1  & F3/5IV/V & 3.40 &$\pm$& 0.03 & Yes & 9.1  & 3.9  &$\pm$& 2.1    \\
 37402 & 45~Myr    & 69.6  & F6V      & 0.82 &$\pm$& 0.03 & No  & -    & \multicolumn{3}{c}{-} \\
 37484 & 45~Myr    & 59.1	& F4V      & 3.49 &$\pm$& 0.02 & Yes & 31   & 18   &$\pm$& 5      \\
 40216 & 45~Myr    & 52.8	& F7V      & 2.38 &$\pm$& 0.02 & No  & -    & \multicolumn{3}{c}{-} \\
 43199 & 45~Myr    & 76.8  & F0III/IV & 4.88 &$\pm$& 0.05 & \ldots & \ldots    & \multicolumn{3}{c}{\ldots} \\
 53842 & 45~Myr    & 57.9	& F5V      & 2.84 &$\pm$& 0.02 & Yes & 1.9  & 93  &$\pm$& 16      \\
207575 & 45~Myr    & 47.0	& F6V      & 2.31 &$\pm$& 0.02 & No  & -    & \multicolumn{3}{c}{-} \\
207964 & 45~Myr    & 46.5	& F0V+F5V  & 9.90 &$\pm$& 0.4  & No  & -    & \multicolumn{3}{c}{-} \\
  \midrule
	3454 & 150~Myr  & 45.4  & F5       & 1.69 &$\pm$& 0.02 & \ldots  & \ldots    & \multicolumn{3}{c}{\ldots} \\
	4277 & 150~Myr  & 52.5  & F8V      & 1.70 &$\pm$& 0.10 & No  & -    & \multicolumn{3}{c}{-} \\
 15407 & 150~Myr  & 49.4  & F5V      & 3.23 &$\pm$& 0.03 & Yes &  430 & 1.01 &$\pm$& 0.35 \\  
 25457 & 150~Myr  & 18.8  & F7V      & 2.05 &$\pm$& 0.01 & Yes & 13.0 & 17   &$\pm$& 4    \\  
 25953 & 150~Myr & 57.0  & F6V      & 1.97 &$\pm$& 0.02 & No  & -    & \multicolumn{3}{c}{-} \\
 31949 & 150~Myr  & 63.1  & F8V      & 1.84 &$\pm$& 0.02 & \ldots  & \ldots    & \multicolumn{3}{c}{\ldots} \\
 61518 & 150~Myr & 61.5  & F5V      & 2.18 &$\pm$& 0.02 & No  & -    & \multicolumn{3}{c}{-} \\
 69051 & 150~Myr & 84.7  & F0III    & 9.27 &$\pm$& 0.09 & \ldots  & \ldots    & \multicolumn{3}{c}{\ldots} \\
103774 & 150~Myr  & 56.5  & F6V      & 3.62 &$\pm$& 0.03 & \ldots  & \ldots    & \multicolumn{3}{c}{\ldots} \\
121560 & 150~Myr  & 24.5  & F6V      & 1.70 &$\pm$& 0.01 & No  & -    & \multicolumn{3}{c}{-} \\
218382 & 150~Myr  & 192   & F8       & 8.10 &$\pm$& 0.20 & \ldots  & \ldots    & \multicolumn{3}{c}{\ldots} \\
219693 & 150~Myr  & 34.1  & F4V      & 5.66 &$\pm$& 0.06 & \ldots  & \ldots    & \multicolumn{3}{c}{\ldots} \\
CD-26 1643 & 150~Myr & 54.8 & F9V   & 1.24 &$\pm$& 0.01 & No  & -    & \multicolumn{3}{c}{-} \\
\bottomrule
 \end{tabular}

\noindent
{\em Notes:}
The data are taken from \cite{zuckerman-et-al-2011, faherty-et-al-2018, gagne-et-al-2018a, gagne-faherty-2018}. The excess emission is given for \textit{Spitzer}/MIPS at 24$\mu$m and/or $70\mu$m. The excess emission for stars with only WISE22 data or upper limits from IRAS is shown as dots.
The fractional luminosities are inferred from a modified blackbody SED model.

\end{table}

Using the studies of members of young moving groups \citep{zuckerman-et-al-2011, faherty-et-al-2018, gagne-et-al-2018a, gagne-faherty-2018}, we identified 29 F~stars in Tab.~\ref{tab:youngstars} for the 45~Myr group and 13 for the 150~Myr group.
For several targets only data up to mid-infrared wavelengths (WISE22) are available or upper limits from IRAS at 25, 60 and 100$\mu$m, but no \textit{Spitzer}/MIPS or other far-infrared data. The presence of far-infrared emission for these targets cannot be ruled out, but none of their SEDs shows excess in the mid-infrared. The detection rates are listed in Tab.~\ref{tab:rates} and given for both the complete samples and the sub-samples only including targets with FIR data.

\begin{table}
\caption{Detection rates for the different samples.
\label{tab:rates}}
\tabcolsep 4pt
\centering
 \begin{tabular}{rccccc}
Sample & $N_\text{Discs}$ & $N_\text{total}$ & $N_\text{FIR}$ & Rate$_\text{total}$ [\%] & Rate$_\text{FIR}$ [\%] \\
\midrule
BPMG   &  9 & 12 & 11 & $75.0^{+25.0}_{-24.5}$  & $81.8^{+18.2}_{-26.8}$  \\
45~Myr  & 11 & 29 & 20 & $37.9^{+15.2}_{-11.3}$  & $55.0^{+22.1}_{-16.3}$  \\
150~Myr  &  2 & 13 &  7 & $15.4^{+20.3}_{-9.9}$   & $28.6^{+37.7}_{-28.5}$  \\
DEBRIS & 21 & 92 & 92 & $22.8^{+6.2}_{-4.9}$\%    & $22.8^{+6.2}_{-4.9}$\%    \\
\bottomrule
 \end{tabular}
 
\noindent
{\em Notes:}
$N_\text{Discs}$ gives the number of disc detections, $N_\text{total}$ is the total number of targets in the sample, $N_\text{FIR}$ is the number of targets with FIR data. The detection rates are given for the complete samples and the sub-samples composed of targets with FIR data assuming the number of disc detections, $N_\text{Discs}$ divided by the sample size. The uncertainties were calculated using the method of \protect{\cite{gehrels-1986}}.
\end{table}

We applied the same modified blackbody model to fit the SEDs of systems in the 45~Myr and 150~Myr groups (see Figs.~\ref{fig:SEDs_THA}and \ref{fig:SEDs_ABDMG}) and inferred stellar parameters, fractional luminosities and blackbody radii using the same method as in \S~\ref{sec:IRExcess} (see Tab.~\ref{tab:youngstars}). 
In the 45~Myr group we find no disc with a blackbody radius below 1~au. Two out of 29 targets possess belts with blackbody radii between 1 and 3~au (6.9\%), five discs lie between 3 and 10~au (17.2\%), two between 10 and 30~au and two between 30 and 100~au (each 6.9\%).
There were only two discs detected within the 150~Myr group. One lies at 1~au the other at 17~au. We note that the disc around 1~au (HD~15407) is only poorly fitted since a strong solid state feature is visible in the SED but that the conclusion of a small blackbody radius is reliable.

Considering NIR, FIR or sub-mm disc radii, only HD~30447 was reported as spatially resolved in scattered light \citep{soummer-et-al-2014} with a detection between 60 and 200~au.

\subsection{Comparing the samples}

\begin{figure}
 \includegraphics[angle = 0, width=\columnwidth]{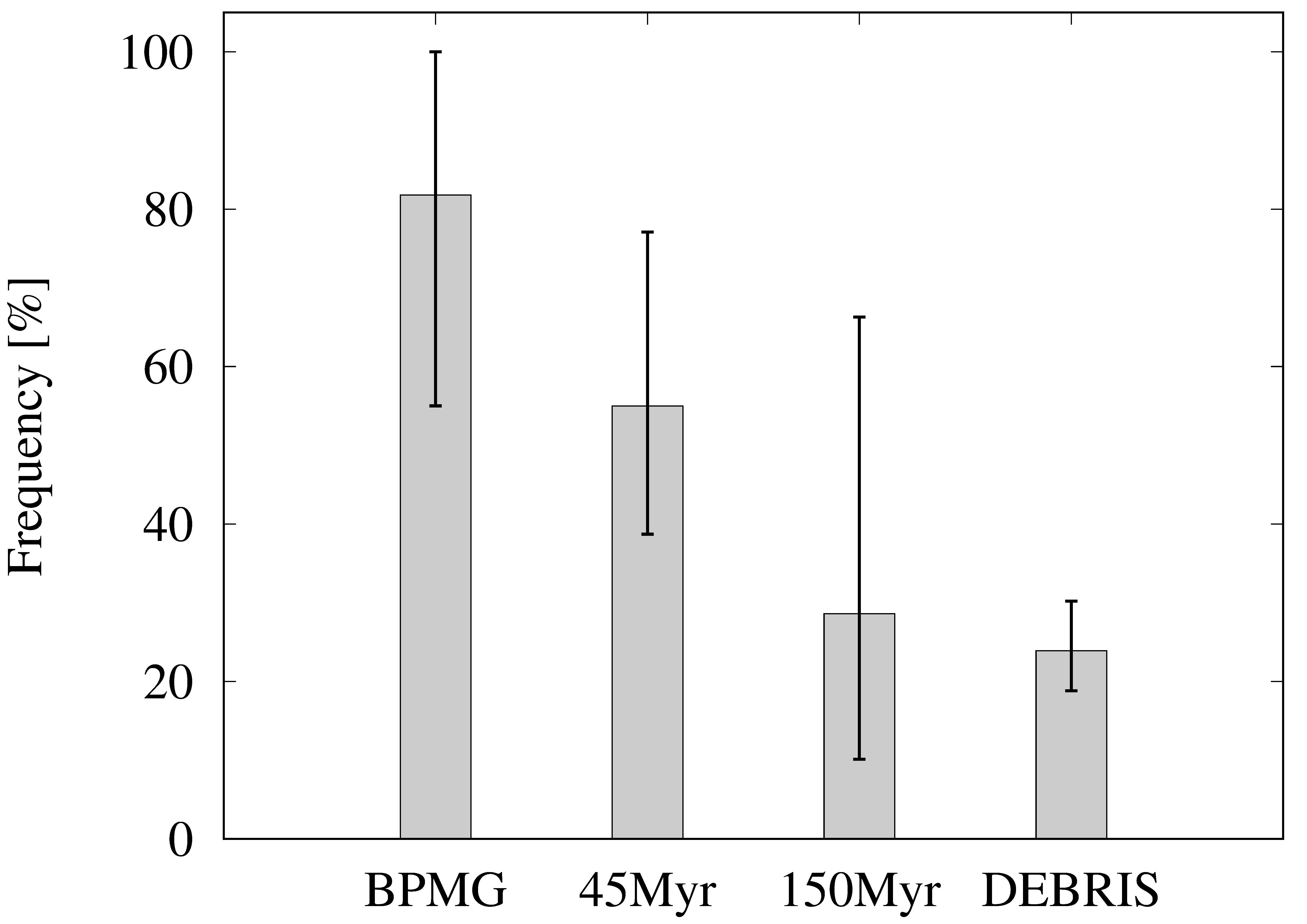}
 \caption[]{Incidence rates for the different samples ordered by age: BPMG (23~Myr), 45~Myr group, 150~Myr group, DEBRIS ($>160$~Myr). The uncertainties are calculated using the method of \protect{\cite{gehrels-1986}}. Only targets with FIR data are taken into account (see Tab.~\ref{tab:rates}) Frequencies are not corrected for completeness.
 \label{fig:Rates}}
\end{figure}

In Fig.~\ref{fig:Rates} we compare the fractions of stars with debris disc detections for each sample which suggests that there might be a decrease of disc frequency with increasing age. 
Using the DEBRIS sample as reference and  Fisher's exact test \citep{fisher-1956} we tested the hypothesis that the incidence rates for the BPMG, the 45~Myr group and the 150~Myr group are similar to the DEBRIS sample. We found that for the BPMG the probability $p = 7.9\times 10^{-4}$, for the 45~Myr group $ p = 0.013$ and for the 150~Myr group $p = 0.68$. The hypothesis is rejected if the $p$-value is smaller than a chosen significance level, $\alpha$ which we set to 0.05. Therefore, we can say that for BPMG and the 45~Myr group the detection rates are not similar to that of the DEBRIS sample. In addition, we tested whether the rates of the BPMG and the 45~Myr group are different from each other and found $p = 0.45$. This means that the BPMG and the 45~Myr group show similar detection rates.
The result leads to the impression that a high frequency of debris discs might be common for F~stars younger than 100~Myr.

\subsection{Fractional luminosity versus radius}
\label{sec:fraction_luminosity}
\begin{figure*}
\includegraphics[width=\textwidth]{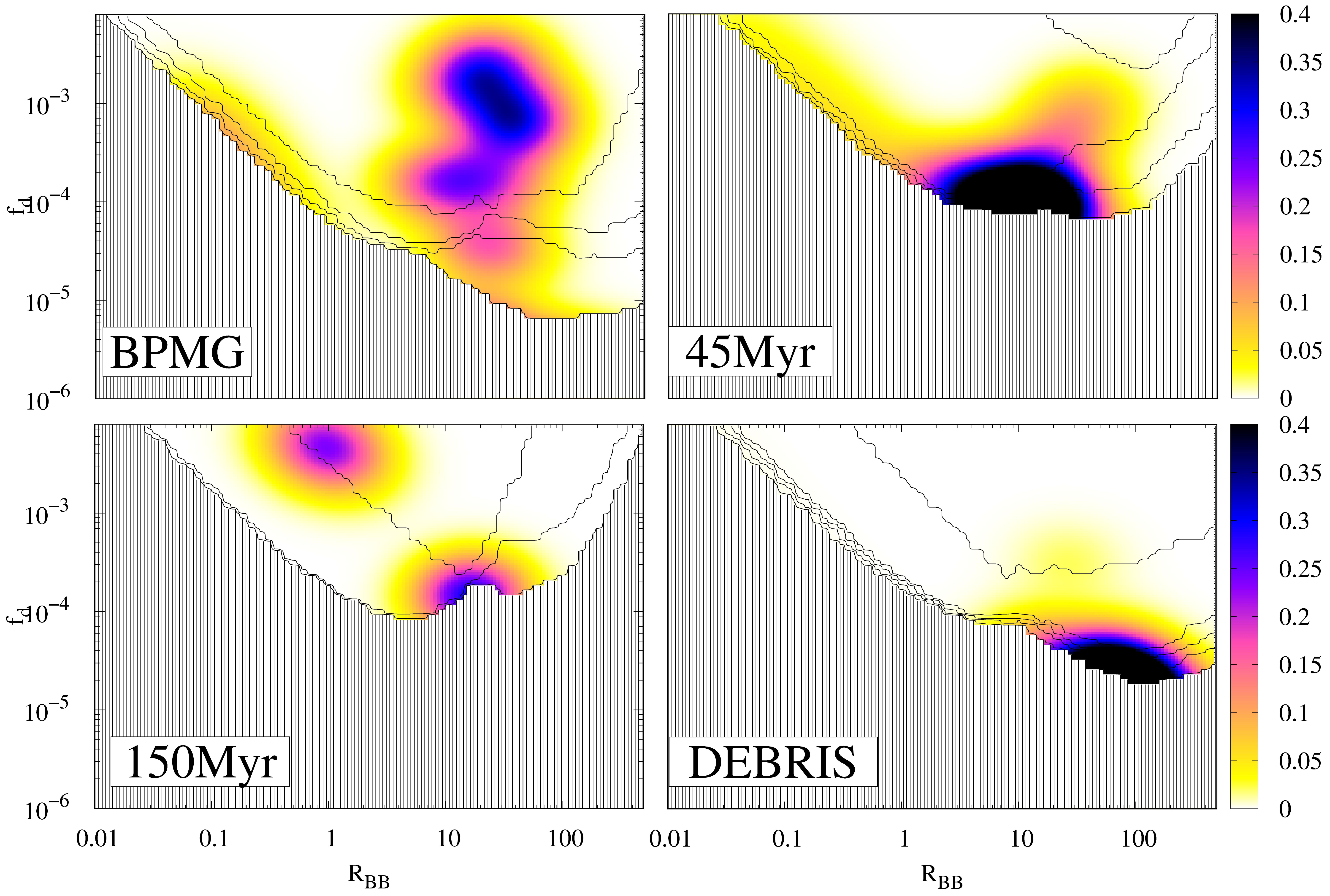}
 \caption{Fractional luminosity as function of blackbody radius for the four samples (BMPG, 45~Myr group, 150~Myr group, and DEBRIS). The colour scale shows the disc incidence, per log(au), per log(unit $f_\text{d}$). The contour lines show the levels of completeness for 0.1, 0.3, 0.5, 0.7 and 1.0 starting from the bottom of the plot.
 \label{fig:fd-R}}
\end{figure*}

Plotting detection rate versus age can be misleading, since different surveys reach different sensitivities to discs, for example due to the different distance of the stars in their samples. 
This sensitivity can be understood within the context of a modified blackbody model, since for each star the region of fractional luminosity vs blackbody radius for which a disc detection would have been possible can be readily quantified. 
Combining this information for all stars in a given sample it is then possible to determine the fraction of stars for which discs could have been detected in different regions of parameter space. This is the basis of the approach taken in Fig.~\ref{fig:fd-R}, which follows on from that used in \cite{sibthorpe-et-al-2018} and \cite{wyatt-2018}.
There we plot the parameter space of fractional luminosity vs blackbody radius for the four samples of F~stars (BPMG, the 45~Myr group, the 150~Myr group, and DEBRIS), noting that the sub-mm disc radius is expected to be $\sim2.5$ times larger than the blackbody radius (\S~\ref{sec:trueradius}). 

To estimate how many discs can be detected in a certain area of parameter space we analysed the targets of each sample independently of whether they were reported to possess a disc or not. Using blackbody radii between 0.01 and 1000~au and fractional luminosities between $10^{-7}$ and $10^{-2}$ we generated a grid of fiducial discs assuming a pure blackbody model. We inferred the flux density of each model disc at wavelengths corresponding to those of observations of each star, e.g. with WISE, \textit{Spitzer}/MIPS, \textit{Herschel}/PACS and ALMA. 
If the total flux density of the fiducial model (star + disc) satisfied 
\begin{equation}
F_\nu > F_\nu^{\text{star}} + 3 \sqrt{\left(\Delta F_\nu^\text{obs} \right)^{2} + \left(\Delta F_\nu^\text{star} \right)^{2} },
\label{eq:detection_limit}
\end{equation}
with $F_\nu^{\text{star}}$ being the flux density of the stellar photosphere, $\Delta F_\nu^{\text{star}}$ its uncertainty and $\Delta F_\nu^{\text{obs}}$ being the uncertainty of the observation, we assumed the model disc to be detected at the wavelength analysed. 
A model disc is counted as a detection as soon as one wavelength band fulfills eq.~(\ref{eq:detection_limit}).
As a result we get the area of parameter space where discs around a certain host star can be detected. For a given sample we calculate the number of stars for which discs could be detected for each node of the grid generating the contour lines shown in Fig.~\ref{fig:fd-R}.
The contour lines are an estimate for the level of completeness of the disc detections. 
For example, if 10 discs are found at a location where discs could have been detected towards 50\% of stars, this suggests that the true number of discs at this location is 10*100/50, since for half of the stars the observations provide no information about the presence of discs at this level.

For the BPMG sample we find that discs with blackbody radii of $\sim20$~au could be detected around 100\% of the stars at fractional luminosities of $1\times10^{-4}$. Discs around  $\sim100$~au could be detected around 10\% of the stars for $f_\text{d} = 6\times10^{-6}$. For the 45~Myr group discs at 100~au could be detected around 10\% of the stars for $f_\text{d}=7\times10^{-5}$ while for the 150~Myr group $f_\text{d} = 2\times10^{-4}$ and for the DEBRIS $f_\text{d}= 3\times10^{-5}$.
The reason for the different sensitivity limits is given by the observations themselves. Some targets have not been studied in detail so that we do not have data longwards of 70$\mu$m 
and only upper limits are available (e.g., from IRAS) which barely constrain the sensitivity limits. 

A second aspect of Fig.~\ref{fig:fd-R} is the actual disc detections. They appear on the plot over the range of $f_\text{d}$ and $R_\text{BB}$ where they could appear according to their likelihood. The likelihood itself was inferred through SED fitting as described in \S~\ref{sec:IRExcess}. 
In Fig.~\ref{fig:fd-R} we use the colour scale to show the fraction of stars for which discs are present. The scale gives the incidence rate of a disc per $\log(\text{au})$, per $\log(\text{unit }f_\text{d})$, per number of targets in the sample. The incidence rate has been corrected for completeness by dividing the observed incidence rates by the sensitivity limits given by the contour lines and was then smoothed with a Gaussian by one order of magnitude in blackbody radius and fractional luminosity. 

Although discs could have been detected down to fractional luminosities of $\sim10^{-6}$ we find that the majority of discs in the BPMG sample is located around $f_\text{d}=10^{-3}$, the area where 100\% of fiducial discs can be detected. 
The 45~Myr group and DEBRIS discs are found in areas closer to the sensitivity limits ($f_\text{d}= 7\times 10^{-5}$ for the 45~Myr group, $f_\text{d}=3\times10^{-5}$ for DEBRIS), some in areas where less than 10\% of the model discs are observable, which results in a higher corrected incidence rate. 
For the 150~Myr group we only have two disc detections, one lying within the area of 100\% completeness the other close to the detection limit. 

Assuming that Fig.~\ref{fig:fd-R} shows comparable debris disc populations at different ages starting from 23~Myr (BPMG) over 45~Myr to older field stars (DEBRIS) we see a decay of fractional luminosity with increasing age which is in agreement with Fig.~\ref{fig:Rates} where we see a decrease in detection rates. While we would expect such a decrease due to collisional evolution it seems that the process takes place in the first 100~Myr (see \S~\ref{sec:samepopulation}). 
Furthermore, the blackbody radii seem to show a slight increase from the BPMG ($\sim30$~au) to DEBRIS ($\sim100$~au). Possible reasons, besides observational biases, will be discussed in \S~\ref{sec:radius_distribution}.

\subsection{Radius distribution}
\label{sec:radius_distribution}
In this section we compare the radii of discs found in the BPMG with those of other young moving groups and field stars. Since most of the targets are not spatially resolved we will look at both blackbody and spatially resolved disc radii to identify possible differences between the samples. 

\subsubsection{Blackbody radii}
\label{sec:rbb}
\begin{figure}
\includegraphics[width=\columnwidth]{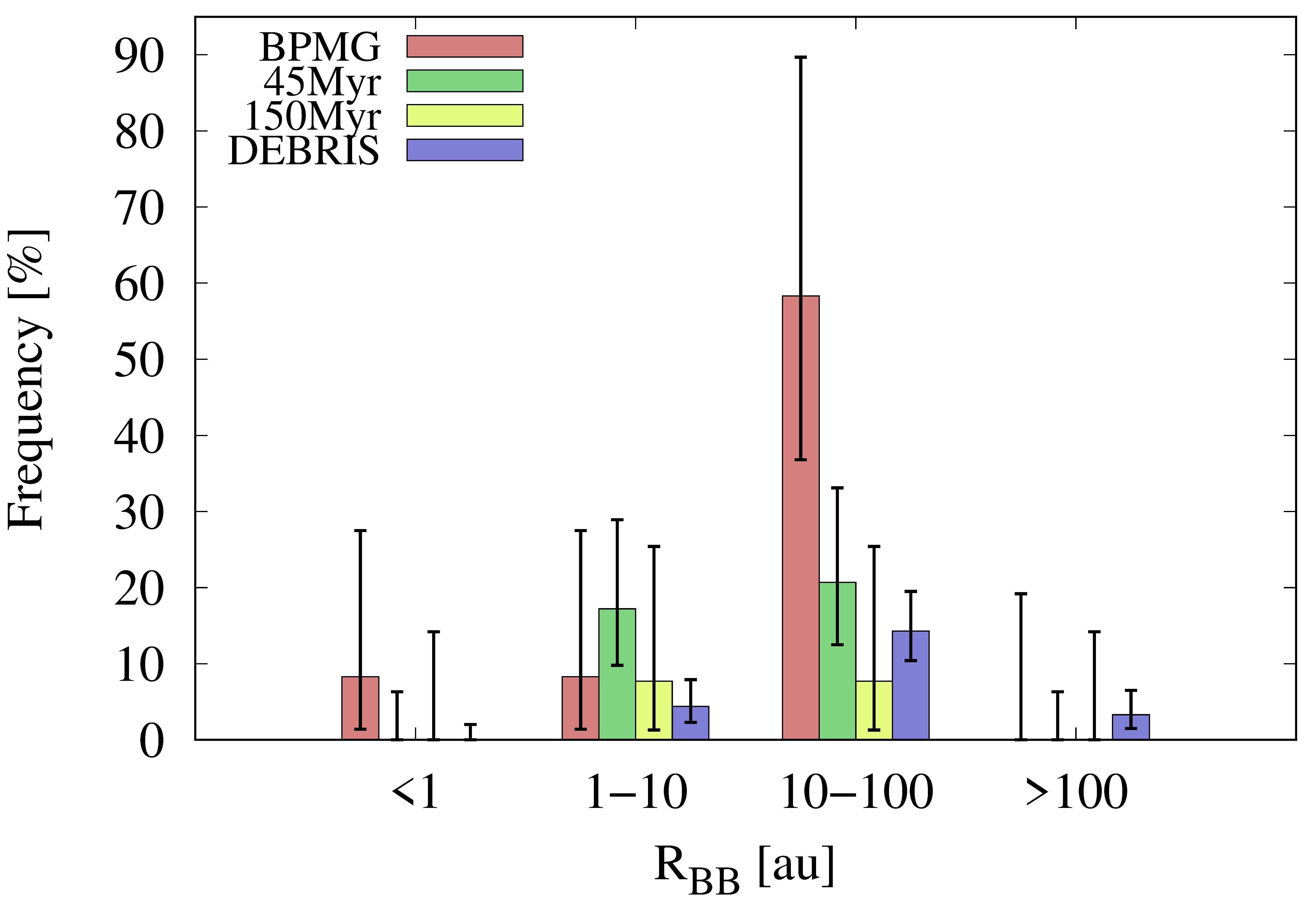}
 \caption[]{Frequency of disc radii for different radius bins assuming the total number of targets in each sample taken from Tab.~\ref{tab:rates}. The uncertainties were calculated using \protect{\cite{gehrels-1986}}.
 \label{fig:radiusdistribution}}
\end{figure}

We focus on the SED results listed in Tabs.~\ref{tab:SEDresults}, \ref{tab:DEBRIS_sample}, and \ref{tab:youngstars} that were used to produce Fig.~\ref{fig:fd-R}.
We compare the blackbody radius of each sample in Fig.~\ref{fig:radiusdistribution} applying four radius bins in logarithmic spacing: $R_\text{BB} < 1$~au comparable to hot dust, $1-10\,\text{au}$ comparable to the warm asteroid belt, $10-100\,\text{au}$ comparable to the cold Kuiper belt, and  $\geqslant 100\,\text{au}$ for larger discs. The frequencies plotted are taken from Tab.~\ref{tab:rates} by comparing the number detected with
the total number of targets in each sample, noting that there could be more discs in each radius bin that are below the detection threshold. 
Most of the discs are found with blackbody radii between 1 and 100~au. For the BPMG sample and the 45~Myr group the majority lies between 10 and 30~au and for the DEBRIS sample between 30 and 100~au.

The latter is the only sample containing discs with blackbody radii larger than 100~au. The DEBRIS sample has a detection limit for large discs down to $f_\text{d}=3\times10^{-5}$ (Fig.~\ref{fig:fd-R}) where the 45~Myr group only shows discs when $f_\text{d}>7\times10^{-5}$, while in the 150~Myr group discs must be brighter than $f_\text{d}=2\times10^{-4}$ to be detected. Thus, it is possible that we miss those large and faint discs in the 45~Myr and 150~Myr group as they would lie below the respective detection limits. 
In the BPMG however, the detection limit lies at $6\times10^{-6}$ and is lower than for the DEBRIS sample. Yet, we did not find any large discs in the BPMG. This might be a result of the low number of targets compared to the DEBRIS sample. 
For example, the probability of detecting one or more $>100$~au disc in the BPMG sample of only twelve stars would be 41.3\% if their incidence rate was the same as that of the DEBRIS sample of 4/92. 

Nevertheless, it seems that the discs in moving groups (BPMG, 45~Myr group) tend to be smaller compared to discs around field stars as seen in DEBRIS (see \S~\ref{sec:fraction_luminosity}).
It could be a systematic increase in physical size with increasing age, or that discs in young moving groups are hotter (and so appear smaller by the $R_\text{BB}$ metric) than around older stars. 
Smaller discs in young moving groups might be expected from collisional theory as those 
could have been depleted around older field stars \citep[see \S~4.2.4 of ][]{wyatt-et-al-2007}.
On the other hand, the discs in the BPMG possess a high fraction of small grains (see \S~\ref{sec:SEDmodelling}) while the particles around comparable field stars are found to be larger \citep{pawellek-krivov-2015}. This might support the idea of hotter discs in young moving groups.  
Nevertheless, the number of targets in each sample is small and the uncertainties are large so that we cannot draw strong conclusions on the difference in the radius distribution. We will consider the influence of collisional evolution in more detail in \S~\ref{sec:detectionrate}.

\subsubsection{Spatially resolved disc radii}
\label{sec:trueradii}

In this section we compare the NIR, FIR, and sub-mm radii inferred from spatially resolved observations from ALMA, \textit{Herschel}/PACS and VLT/SPHERE data. 
Using ALMA, four targets were resolved in the BPMG and two discs (HD~10647 and HD~109085, Tab.~\ref{tab:planetesimals}) in the DEBRIS sample. With \textit{Herschel}/PACS three discs in the BPMG and nine discs in the DEBRIS sample were resolved (Tabs.~\ref{tab:resolveddisc}, \ref{tab:DEBRIS_sample}). 
Considering scattered light observations, four discs in the BPMG were resolved. In the 45~Myr group only HD~30447 was reported as spatially resolved with SPHERE.

\begin{figure}
\includegraphics[width=\columnwidth]{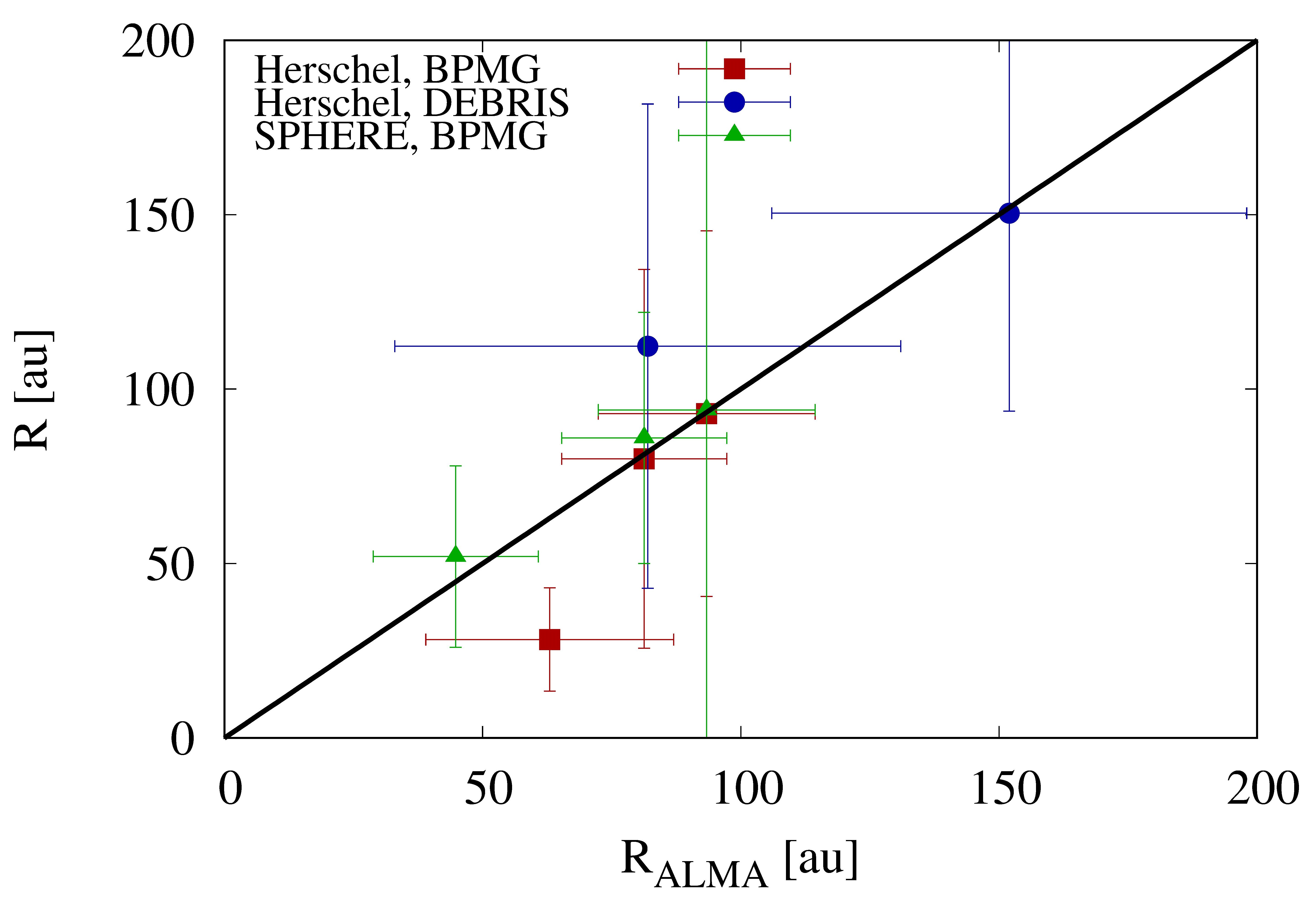}
 \caption[]{Resolved disc radii inferred from \textit{Herschel/PACS} and VLT/SPHERE compared to ALMA radii. The central radius of \textit{Herschel} is assumed to be $0.5\times (R_\text{in, FIR} + R_\text{out, FIR})$. The error bars indicate the disc width inferred from observations.
 \label{fig:radiuscomparison}}
\end{figure}

In Fig.~\ref{fig:radiuscomparison} we compare the ALMA radii to the \textit{Herschel}/PACS and VLT/SPHERE radii for the BPMG and DEBRIS to infer possible biases between the values from the different observations. 
In \S~\ref{sec:herschel} we already found that the \textit{Herschel} and ALMA radii for the BPMG are in good agreement. 
This is also the case for HD~109085 from the DEBRIS sample, while for HD~10647 the \textit{Herschel} radius seems larger compared to ALMA. 
Additionally, SPHERE data show broad extended discs for HD~15115, HD~181327, and HD~191089 with the location of the surface brightness peak being in good agreement with the ALMA radii as well.

Fig.~\ref{fig:radiuscomparison} complements the results found in \cite{pawellek-et-al-2019a}. That study used collisional models and showed that at high resolution the peak of the discs' surface brightness is at the same location in sub-mm and far-infrared images (and is nearly coincident with the planetesimal belt). However, the low surface brightness halo made of small grains that extends beyond the belt gets brighter at shorter wavelengths. 
It is thus possible that due to the halo and the lower resolution of \textit{Herschel} the radii inferred from \textit{Herschel} could appear larger than ALMA radii, which might be the case for HD~10647. 
Based on Fig.~\ref{fig:radiuscomparison} we assume that the disc radii inferred from different telescopes give comparable values.

\begin{figure}
\includegraphics[width=\columnwidth]{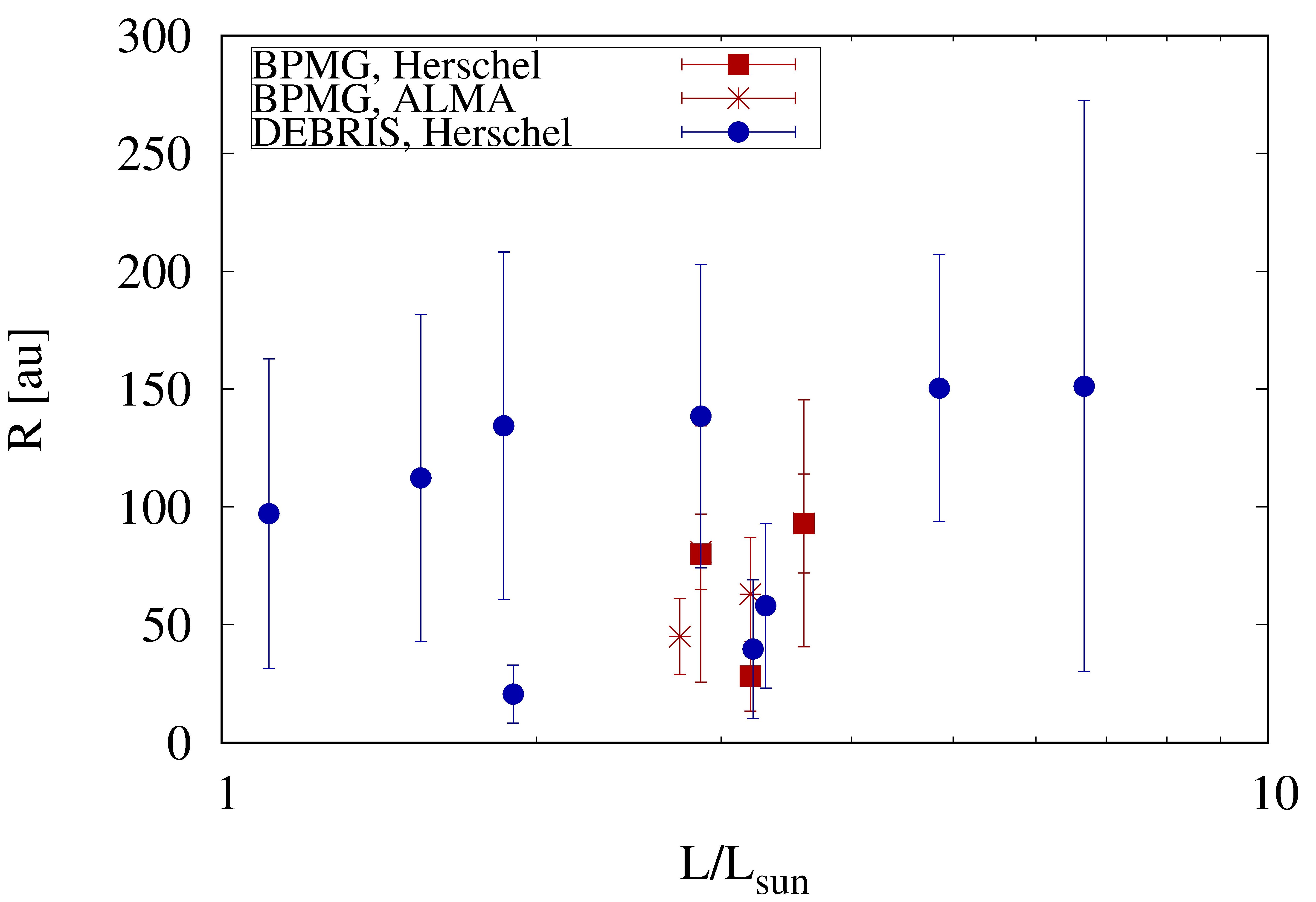}
 \caption[]{Resolved disc radii of the BPMG and DEBRIS as function of stellar luminosity. The radii are inferred from Herschel and ALMA images. The error bars indicate the disc width.
 \label{fig:radius_luminosity}}
\end{figure}

In Fig.~\ref{fig:radius_luminosity} the FIR and sub-mm radii of the BPMG and DEBRIS sample are shown as a function of stellar luminosity with error bars indicating the disc width. 
There are three discs in the DEBRIS sample with FIR radii below 50~au, but the majority of targets (six) possesses radii around $\sim100-150$~au. All discs are found to be broad. Five of the discs with FIR radii between 100 and 150~au also possess blackbody radii larger than 50~au (Tab.~\ref{tab:DEBRIS_sample}) and are in agreement with an expected ratio of sub-mm to blackbody radius ratio of $\sim$2.5 (\S~\ref{sec:trueradius}).
In contrast to the DEBRIS sample the discs in the BPMG are found within 100~au.

Nevertheless, we have to keep in mind that while most of the discs in the DEBRIS sample are larger than the discs in the BPMG the number of spatially resolved discs in both samples is low. Furthermore, seven of the nine resolved discs in DEBRIS were observed with \textit{Herschel}/PACS, but not with ALMA. \textit{Herschel}/PACS has a lower spatial resolution and thus might bias the sample of resolved discs towards larger radii.

\section{Origin of high detection rate}
\label{sec:detectionrate}

We found that the detection rate for debris discs around F-type stars is significantly higher in the BPMG than in the DEBRIS sample. In this section we investigate different scenarios which might explain this phenomenon:\\
(i) The BPMG is representative of the population of stars that become field stars. Hence, the discs seen in both the BPMG and DEBRIS samples are essentially the same population but seen at different ages. This is considered in \S~\ref{sec:samepopulation}.\\
(ii) The BPMG is representative of the population of stars that become field stars. However, the discs seen in BPMG and DEBRIS are \textbf{not} the same population seen at different ages. This is considered in \S~\ref{sec:twopop}.\\
(iii) The BPMG is not representative of the population of stars that become field stars, since the environment of young moving groups is different to that in which field stars formed, and more conducive to the retention of bright discs. This is considered in \S~\ref{sec:diffenv}

\subsection{Same population scenario}
\label{sec:samepopulation}

In the first scenario we assume that the BPMG and DEBRIS samples possess the same population of discs seen at different ages. Therefore, the discs in the BPMG should evolve into discs comparable to the DEBRIS sample. 
To describe the evolution process we use the collisional evolution model (\S~\ref{sec:collisions}) and assume that the disc radius stays constant while the fractional luminosity decreases over time. 

If the largest planetesimals are in collisional equilibrium at the age of the BPMG then the fractional luminosity decreases with $f_\text{d}\propto 1/t$, where $t$ is the time (see eq.~\ref{eq:timescale}). However, it could also have decreased less than this or even stayed constant if the biggest bodies were not yet in collisional equilibrium.

\subsubsection{Modelling detection rates}

\begin{figure}
\includegraphics[angle = 0, width=\columnwidth]{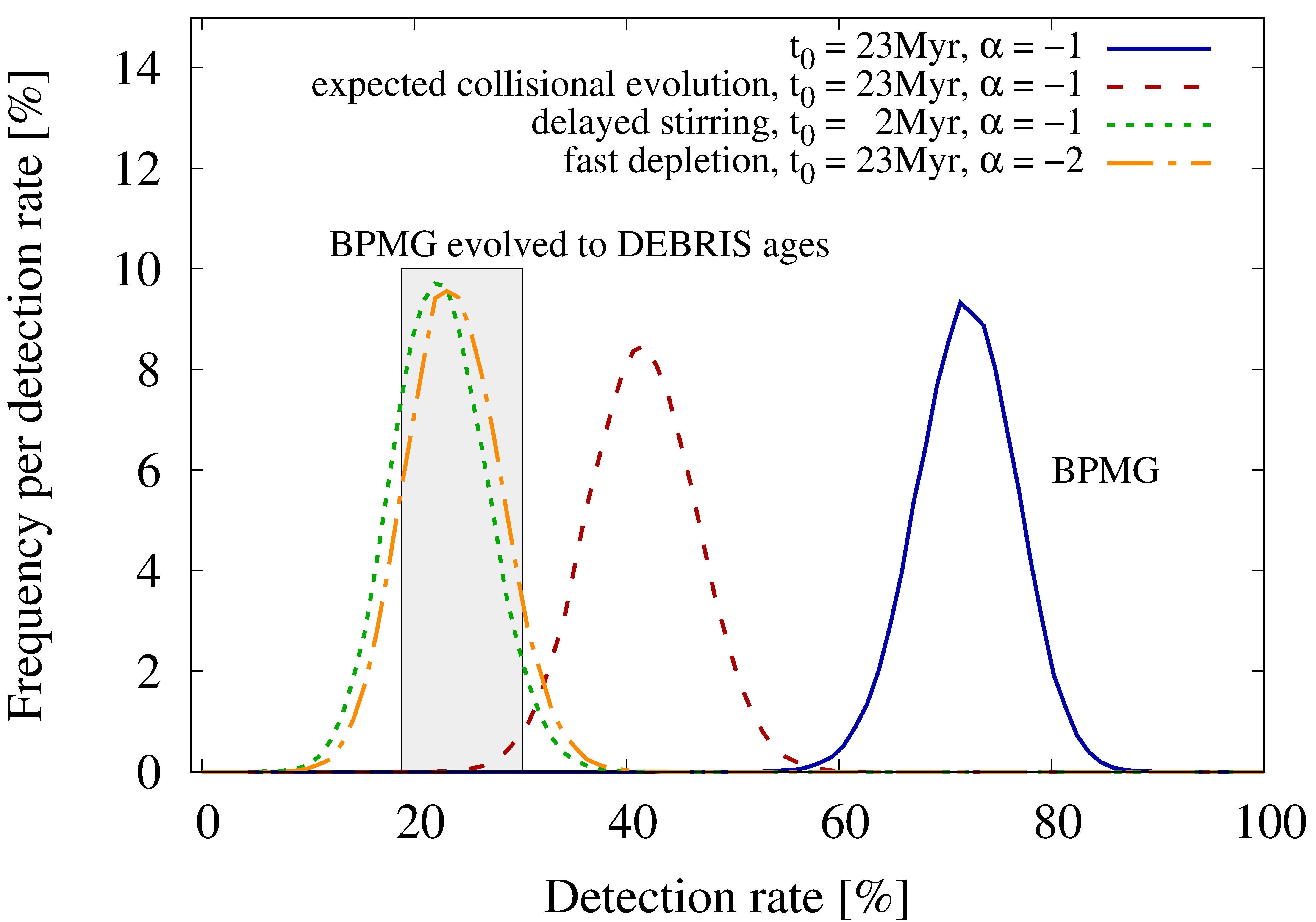}
 \caption[]{Detection rates of a sample made of 92 discs around F-type stars similar to DEBRIS with disc properties similar to the discs in BPMG. The grey region shows the detection rate inferred from the DEBRIS sample \protect{\citep[$23.9^{+6.3}_{-5.1}\%$][]{sibthorpe-et-al-2018}}.
 \label{fig:Evolution}}
\end{figure}

To predict the population that the BPMG would evolve into by DEBRIS ages, we make a generic model. We generate 100,000 artificial samples of 92 targets similar to the size of the DEBRIS sample. Each target is randomly chosen from the 12 systems of the BPMG sample so that each artificial sample is completely made of BPMG targets including those without a disc detection.
We assume that the fractional luminosity follows 
\begin{equation}
f_\text{d}(t) = f_\text{d}(t_0)\,\left(\frac{t}{t_0}\right)^\alpha,
\end{equation}
with $f_\text{d}(t_0)$ being the fractional luminosity at the time $t_0$ and $\alpha$ being a free parameter.
DEBRIS is an unbiased sample of field stars and as such its stars should possess random ages up to the main sequence lifetime. 
We therefore generate random ages ($t$) for the 92 targets in each of our artificial samples and calculate the fractional luminosity $f_\text{d}(t)$ using those inferred from SED modelling of the BPMG sample as values for $f_\text{d}(t_0)$. 
In the next step we consider the parameter space ($R_\text{BB}$, $f_\text{d}(t)$) as shown in Fig.~\ref{fig:fd-R}. 
For each location we know the probability that a star in the sample was observed with sufficient sensitivity to detect the disc. 
We generate a random number between 0 and 1 for each target in the 100,000 samples and compare it to the probability of detecting the disc at its location of parameter space. If the probability is larger than this random number we count the disc as a detection. 
For each of the 100,000 generated samples we assume that the probability to detect a disc at a certain location ($R_\text{BB}$, $f_\text{d}(t)$) is comparable to that of the actually observed DEBRIS sample as shown in Fig.~\ref{fig:fd-R}. 
As a result, we get a distribution of detection rates for the 100,000 artificial samples which is shown in Fig.~\ref{fig:Evolution}. 

To test the model we considered what it would predict for the BPMG, i.e. for a set of 100,000 samples each containing 12 targets randomly taken from the actual BPMG sample with the same age as the BPMG. 
We applied the disc detection probability distribution inferred for BPMG. 
In Fig.~\ref{fig:Evolution} the solid blue line shows the resulting distribution of detection rates peaking around 75\% similar to the actual BPMG sample.

\subsubsection{Expected detection rates}

Given the derived age of the BPMG we set $t_0$ to 23~Myr and $\alpha=-1$ as expected from the collisional evolution model \citep{wyatt-et-al-2007}. 
The dashed red line in Fig.~\ref{fig:Evolution} shows that in this case the DEBRIS sample should have a detection rate of $41\pm10\,\%$ following Poisson statistics with a 95\% confidence level. This is incompatible with the observed detection rate of $22.8^{+6.2}_{-4.9}$\% for the DEBRIS sample. 
Since $\alpha=-1$  is the fastest possible collisional evolution, this suggests that collisional evolution cannot be the explanation if the BPMG and DEBRIS sample possess the same population of discs at different ages.

\subsubsection{Delayed stirring}

The observed detection rate of the DEBRIS sample could be explained if the cascade was initiated only recently, i.e., if the collision age was closer to $t_0\sim2$~Myr instead of the stellar age of 23~Myr which is shown with the dotted green line in Fig.~\ref{fig:Evolution}.

However, this is not realistic since protoplanetary discs dissipate within a few Myr \citep[e.g.,][]{pascucci-et-al-2006}. While delayed stirring is expected in some models \citep[e.g.,][]{kenyon-bromley-2008}, this explanation would require all discs to wait 21~Myr before being ignited, whereas all discs are located at different radii and so should have different evolution timescales.

\subsubsection{Fast depletion}

The other explanation that is compatible with the assumption of similar disc populations in the BPMG and DEBRIS samples is that the evolution is faster than $t^{-1}$. The dash-dotted orange line in Fig.~\ref{fig:Evolution} shows that $t^{-2}$ is in agreement with this hypothesis.
However, such rapid evolution is no longer consistent with simple collisional evolution models. 
Collisional models might still be able to reproduce rapid evolution of disc luminosity if in addition to depletion of the large bodies there is also a change in the equilibrium dust size distribution, e.g., an over-abundance of small grains above steady state at BPMG ages. 

There might be physical motivation for the dust size distribution to evolve, say if the quantity of sub-blowout grains destroying the larger particles is changing or if there is gas preventing small dust being removed at young ages. Indeed, the collision model introduced by \cite{loehne-et-al-2017} shows an increase in particles slightly larger than the blow-out size \citep[e.g., Fig.~6 in][]{loehne-et-al-2017}.
However, there is no evidence for a change in small dust properties in the SED models' $s_\text{min}$ or $q$, and there is not significant gas in these systems. The only target with a gas detection is HD~181327 (see \S~\ref{sec:gas}) for which \cite{marino-et-al-2016} found only low gas masses making the presence of gas unlikely to influence the dust in this disc. Hence, gas is also unlikely to explain the fast depletion of the whole sample. This leads to the conclusion that something other than collisions is depleting the BPMG discs.

One potential problem with this scenario is that the evolution with $t^{-2}$ suggested by our model is not compatible with other studies which have shown that the discs of Sun-like stars evolve slowly on the main sequence. While ages are hard to determine for those stars, there seems to be slow evolution ($\sim t^{-0.5}$) beyond a few 100~Myr \citep[e.g.,][]{trilling-et-al-2008, holland-et-al-2017}. However, the $t^{-2}$ trend only represents an average evolution.
A solution to this problem is thus that there is a process that depletes the BPMG discs that acts even faster than $t^{-2}$, but only on timescales of order 100~Myr 
following which the slower collisional depletion resumes for any remaining planetesimals that go on to supply the dust seen in the DEBRIS population. 

The Solar System's Edgeworth-Kuiper belt underwent depletion by two orders of magnitude in mass on a timescale of 10-100~Myr as a result of the dynamical instability in the planetary system, so this is one possibility \citep[][]{gomes-et-al-2005}. 
Others could be embedded planetary embryos, or planets that migrate into the discs \citep[e.g.,][]{gomes-et-al-2005, levison-et-al-2008, izidoro-et-al-2014, nesvorny-et-al-2015}. 
Given the depletion timescale inferred above, we use eq.~(3) of \citet{shannon-et-al-2016} to estimate the mass of potential planetary perturbers by setting the timescale of such perturbers to clear their surrounding area from dust to 100~Myr. We find masses between 20 and 170 earth masses (corresponding to 1 and 10 Neptune masses). 
With currently available telescopes such planets could very well be undetected within the discs in the BPMG. 
Based on results from different observational instruments like \textit{Kepler} or HARPS summarised in the exoplanet database\footnote{http://exoplanet.eu/} \citep{schneider-2011}, $\sim$260 close-in planets were detected around F-type stars by radial velocity or transit observations, $\sim50$ of them possessing masses between 1 and 10 Neptune masses. This is just an estimate, since in many cases the total masses of the planets could not be inferred. Furthermore, most of the planets detected are located close to the host star and are far away from the typical debris discs investigated in our study. \cite{suzuki-et-al-2016} analysed data from the Microlensing Observations in Astrophysics (MOA) and estimates that cold Neptunes located beyond the snow line of stellar systems and thus, closer to debris discs might be more common than their close-in hot siblings.

51~Eri is the only system with a planet detection in our sample. Due to a lack of spatially resolved images of the disc, we cannot rule out that the Jupiter-mass planet might be located close to the disc and thus, accelerating its depletion. An indicator for this scenario could be the already low disc's fractional luminosity of $5\times10^{-6}$. Indeed, a Jupiter-mass planet located within our BPMG discs would only need $\sim20$~Myr to clear its surrounding area \citep[eq.~3, ][]{shannon-et-al-2016}.

Another possible depletion mechanism is the disintegration of planetesimals due to heating. \cite{lisse-et-al-2020} showed that high energy stellar flares are able to heat dust in close-in debris discs to temperatures of $\sim1000$~K. According to studies of Kepler/K2 stars \citep[][]{davenport-2016, vandoorsselaere-et-al-2017}, 1.6\% of young F-type stars show such flares.
However, the majority of the discs in the BPMG lie too far out ($\sim80$~au) for this to be important. The close-in disc around HD~199143 might be a candidate for such a depletion mechanism though collisions would be expected to deplete this disc by DEBRIS ages \citep{wyatt-et-al-2007}

Planetesimals could also be depleted by stellar radiation forces such as the YORP effect. 
However, the relatively slow evolution of Solar system asteroids \citep{durech-et-al-2018}
suggests that this radiation effect might be negligible, since it should be weaker for planetesimals at several tens of au.

\subsection{Two population scenario}
\label{sec:twopop}

In \S~\ref{sec:samepopulation} we assumed that the discs seen in the BPMG and DEBRIS samples are the same population seen at different ages.
While it is possible that the BPMG is representative of the population of stars that become field stars, comparable to the DEBRIS sample, it is also possible that the BPMG and DEBRIS samples do not show the same population of discs. 

This could be the case for example if the discs in the DEBRIS sample are belts of planetesimals like the Edgeworth-Kuiper belt that formed as part of the planet formation process,
whereas those in BPMG are remnants of the primordial dust that is swept up into a ring 
by the depleting gas that forms planetesimals by streaming instability 
\citep[e.g.,][]{johansen-et-al-2007, johansen-et-al-2012, carrera-et-al-2015, carrera-et-al-2017, schaffer-et-al-2018}. 
If that is the case the implication is that BPMG stars would have two belts - the bright primordial dust belt and the DEBRIS-like belt with large planetesimals which may be too faint to detect at this young age. This scenario may be supported by the tentative outer belt found around HD~181327 \citep[e.g.,][]{marino-et-al-2017}. 

However, the planetesimals in the two proposed belts in this scenario would deplete by collisions and thus by $t^{-1}$ as predicted by collisional models which is thus incompatible with the rate of $t^{-2}$ seen in Fig.~\ref{fig:Evolution}.  
Nevertheless, a key difference between the two populations might be that the DEBRIS belts would be planetesimals formed in stable regions of the planetary system (like the Edgeworth-Kuiper belt and Asteroid belt), whereas the BPMG belts could be deposited anywhere, since this would depend on how the gas disc depletes which could be driven by photo-evaporation processes. Thus, unstable regions liable to dynamical depletion as discussed in \S~\ref{sec:samepopulation} may be more likely for such BPMG belts.

Another possibility is that the ``planetesimals'' formed in these unstable regions could be more loosely bound and liable to disruption.
To consider this, we compare the minimum sizes of planetesimals inferred in \S~\ref{sec:planetesimalsizes} of discs in both the BPMG and DEBRIS samples. Despite the large variation of the sizes, and the small number of discs being compared, we see a similar range between both samples.
This does not support the hypothesis of two belts with different planetesimal properties, but it should be noted that if planetesimals form differently then they may have different $Q_\text{D}^*$ and so different planetesimal sizes would be inferred. Nevertheless, the small planetesimal sizes are in agreement with recent studies suggesting the absence of large planetesimals \citep{krivov-et-al-2020}.

Considering the disc radii inferred from spatially resolved images (\S~\ref{sec:imaging}) the radii of discs in the BPMG lie between 45 and 94~au. 
We might expect systematic differences in the disc radius between the BPMG and DEBRIS samples for this scenario. This might be supported by the fact that the spatially resolved discs in DEBRIS tend to be larger than those in the BPMG (see \S~\ref{sec:trueradii}).
However, while the bright belts seen in the BPMG should be depleted and only the fainter DEBRIS belts remain detectable at DEBRIS ages both planetesimals rings should be present at BPMG ages.

We investigated the possibility of detecting a DEBRIS-like outer planetesimal belt around a BPMG star with a bright inner belt. We took HD~181327 with its bright ring at 80~au and considered an additional fainter outer belt at 150~au with a width of 46~au similar to that of HD~109085 from the DEBRIS sample.
Originally, HD~109085 was observed with ALMA at 880$\mu$m and a sensitivity of 30$\mu$Jy/beam \citep{marino-et-al-2017} while HD~181327 was observed at a higher spatial resolution and a sensitivity of 27$\mu$Jy/beam (see Tab.~\ref{tab:resolveddisc}). 
The surface brightness of HD~109085 is  $\sim200\mu$Jy/beam \citep[Fig.~1 in ][]{marino-et-al-2017}. If the disc was located around HD~181327 and observed with the sensitivity of 27$\mu$Jy/beam we would detect it at a 0.15$\sigma$ level (in each beam). With azimuthal and radial averaging the detection would reach a $3.4\sigma$ level with ALMA band 7 for the whole disc. Applying an observational setting like that in \cite{marino-et-al-2016} with a lower spatial resolution we would even reach a level of $\sim18\sigma$. 

We do not detect such outer discs in the BPMG (the only exception being HD~181327 for which there is a tentative detection at $\sim200$~au). This could either mean those outer belts do not exist or they are too faint to detect at that age (e.g., because the collisional cascade has yet to be fully initiated).

\subsection{Different star formation environments}
\label{sec:diffenv}

The third scenario to explain the higher detection rate supposes that the environment of young moving groups is different to that in which field stars form, such that these regions might be more conducive to the retention of bright discs.

\cite{deacon-et-al-2020} analysed the binary fraction of open clusters such as the Pleiades and compared it to less dense associations including the BPMG. That study found that the rate of wide multiples (between 300 and 3000~au) is higher in young moving groups (14.6\%) than in field stars (7.8\%) or open clusters (Hyades, 2.5\%) which is in agreement with our results 
(see \S~\ref{sec:multiplicity}). \cite{deacon-et-al-2020} concluded that the rate of multiple systems might be influenced more strongly by environmental factors than by age which supports the idea of different formation environments for young moving groups and field stars. It seems that wide separation multiple systems are more effectively formed in less dense regions such as young moving groups. However, as shown by \cite{yelverton-et-al-2019}, an influence of wide-separation binaries on the detection rate of debris discs could not be found so far (\S~\ref{sec:multiplicity}). 

Of greater importance might be the evolution of multiple stellar systems.  \cite{reipurth-mikkola-2012} and \cite{elliott-bayo-2016} suggested that the fraction of such systems might decrease over time as the stellar systems become unstable and break-up within $\sim100~$Myr. 
It is possible that firstly, the break-ups destroy the debris discs in the system, and secondly the break-ups lead to a higher rate of stellar flybys in the moving group which truncate and/or deplete the debris discs. As a result, the radii of the discs in the field might be on average smaller and fainter than those in the BPMG. 

This idea however is not supported by our results on the radial distribution in the BPMG and DEBRIS where the discs in DEBRIS tend to be larger than in the BPMG (see \S~\ref{sec:trueradii}). 
\cite{lestrade-et-al-2011} investigated the depletion of debris discs due to flybys during the first 100~Myr and found that only high-density regions like Orion with star densities $>20,000$~pc$^{-3}$ have a significant impact on the discs.
Similarly, \cite{vincke-et-al-2018} analysed the impact of the high-density environment found in open clusters, such as Trumpler~14, on discs and planetary systems. That study found that during the initial phase of evolution stellar flybys resulted in $\sim90\%$ of discs having a 
radial extent smaller than 50~au. For $\sim47\%$ of the discs the radii were even smaller than 10~au. At later evolution stages of the clusters the discs were barely influenced by stellar interactions. 
Assuming that field stars formed in dense clusters \citep{eggen-1958} it is possible that stellar flybys truncated a number of their protoplanetary discs leading to a smaller fraction of debris discs with large radii and/or a lower incidence of debris discs around field stars \citep[][]{hands-et-al-2019}. Again, this is not supported by the radii of spatially resolved discs (see \S~\ref{sec:trueradii}), but since we analysed only a small number of them around field stars these might be the ones which were not altered by stellar flybys. 

Nevertheless, it might be possible that the detection rate of debris discs around field stars is low from an early phase, since their protoplanetary predecessors were already truncated. In contrast, the discs which formed in less dense regions like the BPMG might retain their high detection rates since stellar flybys are less frequent. This might be observationally testable by comparing disc incidences in \S~\ref{sec:comparison} for different clusters with those of more dense clusters at a comparable age.  
Recently, \cite{miret-roig-et-al-2020} derived a disc detection rate of $9\pm9$\%  for stars ranging from F5 to K5 in the 30~Myr old cluster IC~4665 based on Spitzer and WISE data. This is much lower compared to the rates we find for the BPMG and the 45~Myr group (75\% and 38\%, Tab.~\ref{tab:rates}). However, we note that the cluster has a distance of 350~pc in contrast to the close-by targets analysed in our study so that many discs might be undetected. A more detailed analysis for example repeating the analyses of the $f_\text{d}$ vs $R_\text{BB}$ parameter space like \S~\ref{sec:fraction_luminosity} for IC~4665 would be needed to draw reliable conclusions on this scenario.

\section{Conclusions}
\label{sec:conclusions}

In the first part of this study we analysed a sample of twelve F-type stars in the BPMG and investigated different properties of the systems. In the second part we compared the results of the BPMG to those of other samples of young moving groups and field stars to analyse possible disc evolution processes. 

We found that nine stars in the BPMG possess debris discs leading to a detection rate of 75\%. This is significantly higher than found in unbiased samples of field stars where only $\sim$23\% of the targets show evidence for debris discs \citep{sibthorpe-et-al-2018}.

Five out of the nine discs were spatially resolved with either ALMA or VLT/SPHERE allowing us to study their radial and grain size distribution in more detail. The disc around HD~164249 was spatially resolved with ALMA for the first time. 
The disc radii lie between 45 and 94~au and are comparable to the radii found for other debris discs and protoplanetary discs, but tend to be slightly smaller compared to spatially resolved discs found in the DEBRIS sample of field stars.

We compared the disc radius to blackbody radius ratio derived from SED modelling to the relation based on \textit{Herschel} data presented in \cite{pawellek-krivov-2015} and found that the resolved discs in the BPMG possess smaller radii than expected. Since ALMA has a higher spatial resolution than \textit{Herschel} we inferred the sub-mm disc to blackbody radius ratio - stellar luminosity relation from a sample of ALMA data \citep[][]{matra-et-al-2018}. The resulting relation shows a weaker decrease of the radius ratio with increasing stellar luminosity. 

The minimum grain sizes of the SED models are in agreement with the blow-out grain sizes of the discs as we would expect from collisional evolution. The exception is HD~15115 with an $s_\text{min}$ of $\sim$5$\mu$m which is also the only disc showing the presence of a warm inner component.
This result is somewhat different to earlier studies \citep{pawellek-et-al-2014} which found an average size of 5$\mu$m for a sample of 34 discs. A reason might be that 66\% of those discs were fitted with a warm inner component, but nevertheless, the small $s_\text{min}/s_\text{blow}$ ratio indicates that the discs are collisionally very active with high levels of dynamical excitation. However, a more detailed analysis is needed to draw strong conclusions.

We compared the sample of BPMG stars to other young moving groups and old field stars, finding that the detection rate of debris discs is significantly higher in young moving groups than in the field star sample. 
Furthermore, the discs in the BPMG possess a higher fractional luminosity. From collisional evolution models we would expect the same discs around older stars to be fainter, which might also cause a lower detection rate. 
However, applying those models we found evolving the BPMG sample to DEBRIS ages 
results in a population with significantly higher detection rate than that observed for the actual DEBRIS sample. We investigated different scenarios explaining this. 

In the first scenario we assumed that the BPMG and the DEBRIS samples show the same disc population at different ages. We found that the observed detection rate could be explained by a delayed ignition of the collisional cascade, but that this option seems unlikely since all discs would need to be delayed by the same $\sim20$~Myr timescale.
A more likely scenario is that additional depleting processes are at work so that the disc evolution cannot be explained by collisional processes alone. 
Depletion through gravitational interaction with unseen planets is one possibility. We found that Neptune-sized planets orbiting within discs can cause depletion on the required $\sim100$\,Myr timescales, and are small enough to remain undetected in current observations. 
For discs close to the star high energy stellar flares and other radiation effects (e.g., YORP) are also possible but less likely.
Whatever the processes are they have to work between 10 and 100~Myr since previous studies showed that disc evolution is slower at older ages \citep[e.g.,][]{holland-et-al-2017}. 

The second scenario assumed that the discs in young moving groups and around old field stars are not part of the same population. It is possible that discs in the BPMG possess two belts, one made of large planetesimals formed by planet formation processes comparable to the Edgeworth-Kuiper belt, and another made of remnants of the primordial dust that grow to planetesimal sizes during the disc dispersal process. This might be supported by the different radii found for the BPMG and DEBRIS samples, but since we studied only a small number of discs, the actual radius distribution is not well characterised yet. However, while the two-population scenario is not impossible, we would still need to invoke a rapid depletion as proposed in the first scenario (\S~\ref{sec:samepopulation}).

In the third scenario we assumed that the birth environment of stars is different for young moving groups and field stars so that their respective discs might be different as well. The influence of stellar flybys in circumstellar discs is significant at early stages of the evolution for dense stellar clusters like Orion \citep[e.g.,][]{lestrade-et-al-2011, vincke-et-al-2018}, but barely contributes to the depletion of debris discs found in less dense associations like the BPMG. 
On the other hand, field stars are supposed to form in regions of higher stellar density so that stellar flybys might truncate the discs at an early evolutionary stage. Therefore, a large fraction of discs around field stars might possess a radius too small to be detected while discs with larger radii in moving groups remain detectable. This is not supported by the radii of spatially resolved discs in the BPMG and DEBRIS, but it is possible that we only see those discs around field stars that were not truncated.
A possibility to test this hypothesis is to analyse the detection rates of debris discs in young dense clusters. Indeed, studies found lower disc detections for the clusters \citep[e.g. IC~4665,][]{miret-roig-et-al-2020}, but this might be biased by the large distance of IC~4665 rather than an actual difference in the fraction of stars with discs.

\section*{Acknowledgements}
NP thanks Alexander Krivov and Torsten L\"ohne for fruitful discussions.
GMK was supported by the Royal Society as a Royal Society University Research Fellow.

The Combined Atlas of Sources with Spitzer/IRS Spectra (CASSIS) is a product of the 
Infrared Science Center at Cornell University, supported by NASA and JPL.

ALMA is a partnership of ESO (representing its member states), NSF (USA) and NINS (Japan), together with NRC (Canada), MOST and ASIAA (Taiwan), and KASI (Republic of Korea), in cooperation with the Republic of Chile. The Joint ALMA Observatory is operated by ESO, AUI/NRAO and NAOJ.

\section*{Data availability}
The data underlying this article will be shared on request to the corresponding author. 
The ALMA and \textit{Herschel} data are publicly available and can be queried and downloaded directly from the ALMA archive at https://almascience.nrao.edu/asax/ and from the \textit{Herschel} archive at
http://archives.esac.esa.int/hsa/whsa/.





\newcommand{\AAp}      {A\& A}
\newcommand{\AApR}     {Astron. Astrophys. Rev.}
\newcommand{\AApS}     {AApS}
\newcommand{\AApSS}    {AApSS}
\newcommand{\AApT}     {Astron. Astrophys. Trans.}
\newcommand{\AdvSR}    {Adv. Space Res.}
\newcommand{\AJ}       {AJ}
\newcommand{\AN}       {AN}
\newcommand{\AO}       {App. Optics}
\newcommand{\ApJ}      {ApJ}
\newcommand{\ApJL}     {ApJL}
\newcommand{\ApJS}     {ApJS}
\newcommand{\ApSS}     {Astrophys. Space Sci.}
\newcommand{\ARAA}     {ARA\& A}
\newcommand{\ARevEPS}  {Ann. Rev. Earth Planet. Sci.}
\newcommand{\BAAS}     {BAAS}
\newcommand{\CelMech}  {Celest. Mech. Dynam. Astron.}
\newcommand{\EMP}      {Earth, Moon and Planets}
\newcommand{\EPS}      {Earth, Planets and Space}
\newcommand{\GRL}      {Geophys. Res. Lett.}
\newcommand{\JGR}      {J. Geophys. Res.}
\newcommand{\JOSAA}    {J. Opt. Soc. Am. A}
\newcommand{\MemSAI}   {Mem. Societa Astronomica Italiana}
\newcommand{\MNRAS}    {MNRAS}
\newcommand{\PASJ}     {PASJ}
\newcommand{\PASP}     {PASP}
\newcommand{\PSS}      {Planet. Space Sci.}
\newcommand{\RAA}      {Research in Astron. Astrophys.}
\newcommand{\SolPhys}  {Sol. Phys.}
\newcommand{\SolSysRes}{Sol. Sys. Res.}
\newcommand{\SSR}      {Space Sci. Rev.}

\bibliographystyle{mnras}
\bibliography{Paper.bbl}


\appendix

\section{SEDs of debris discs around nearby F stars}

\subsection{45~Myr group}
\begin{figure*}
\centering
\includegraphics[width=0.9\textwidth]{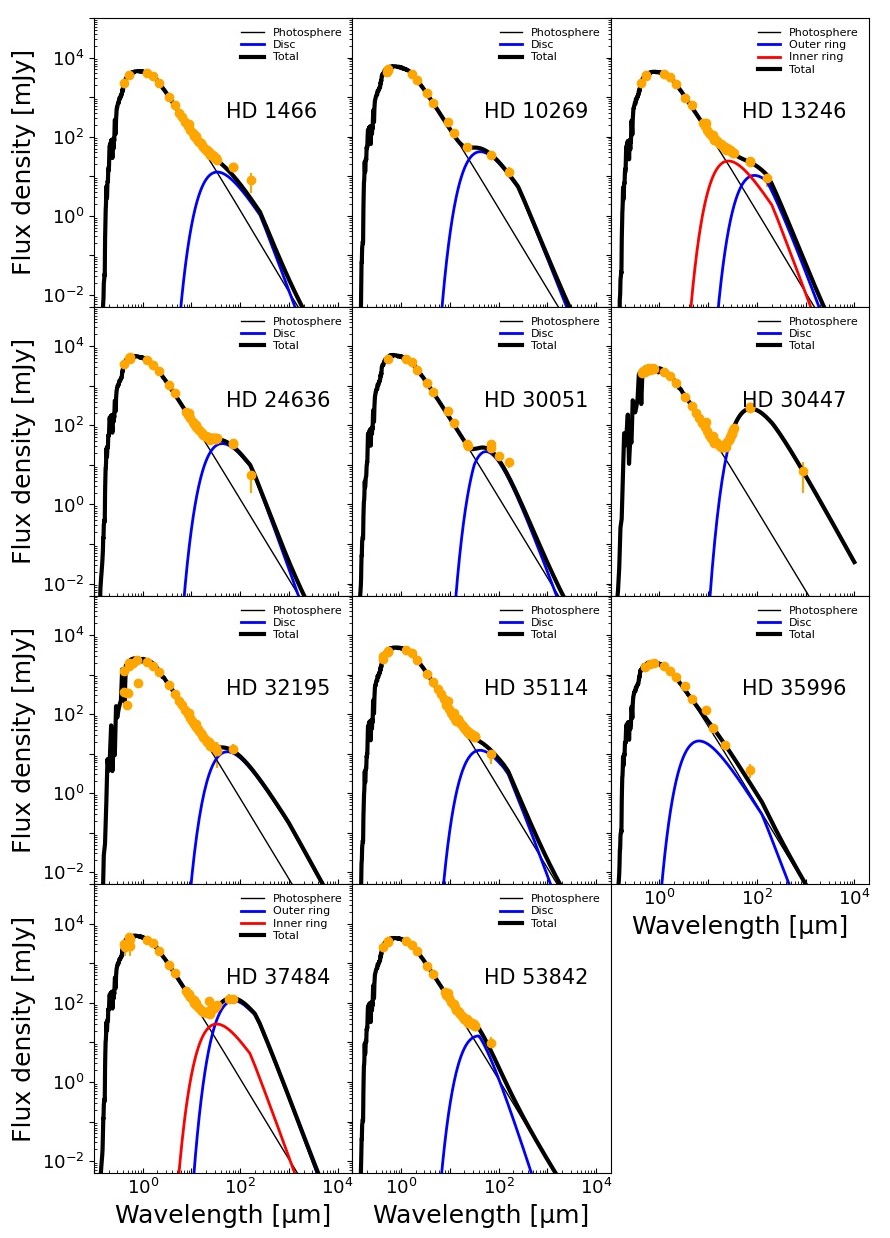}
 \caption{SEDs for the debris discs detected around F~stars in the 45~Myr group.  
 \label{fig:SEDs_THA}}
\end{figure*}

We analysed the sample of 29 F-type stars found in the 45~Myr group (see \S~\ref{sec:youngMG}) and found that eleven of them exhibit significant mid-infrared excess. 
We fitted the SEDs of these targets with a modified blackbody model which is described in detail in Section~\ref{sec:IRExcess}. The results are shown in Fig.~\ref{fig:SEDs_THA}. 

\subsection{150~Myr group}
\begin{figure*}
\includegraphics[width=1.2\columnwidth, height = 6cm]{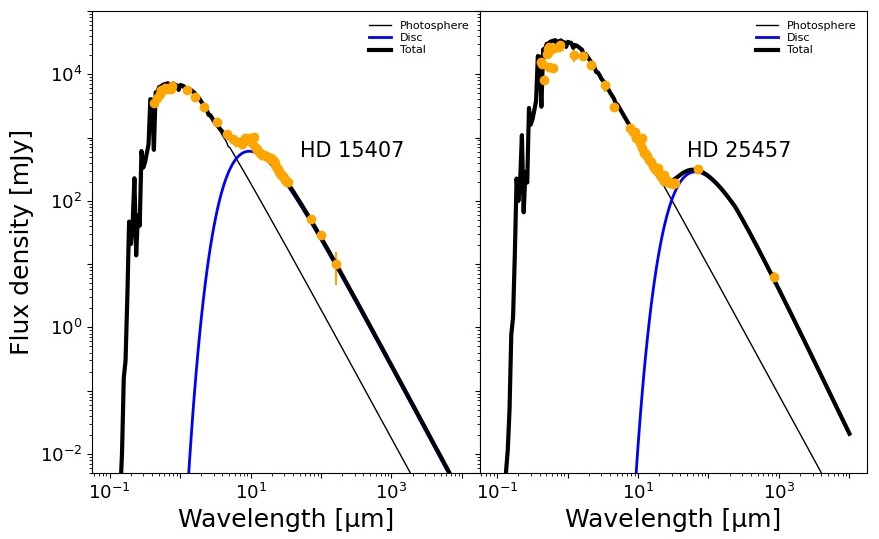}
 \caption{SEDs for the debris discs detected around F~stars in the 150~Myr group.  
 \label{fig:SEDs_ABDMG}}
\end{figure*}

We analysed the sample of 13 F-type stars found in the 150~Myr group (see \S~\ref{sec:youngMG}) and found that two of them exhibit significant mid-infrared excess. 
We fitted the SEDs of these targets with a modified blackbody model which is described in detail in Section~\ref{sec:IRExcess}. The results are shown in Fig.~\ref{fig:SEDs_ABDMG}. 

\subsection{DEBRIS}
\begin{figure*}
\includegraphics[width=0.9\textwidth]{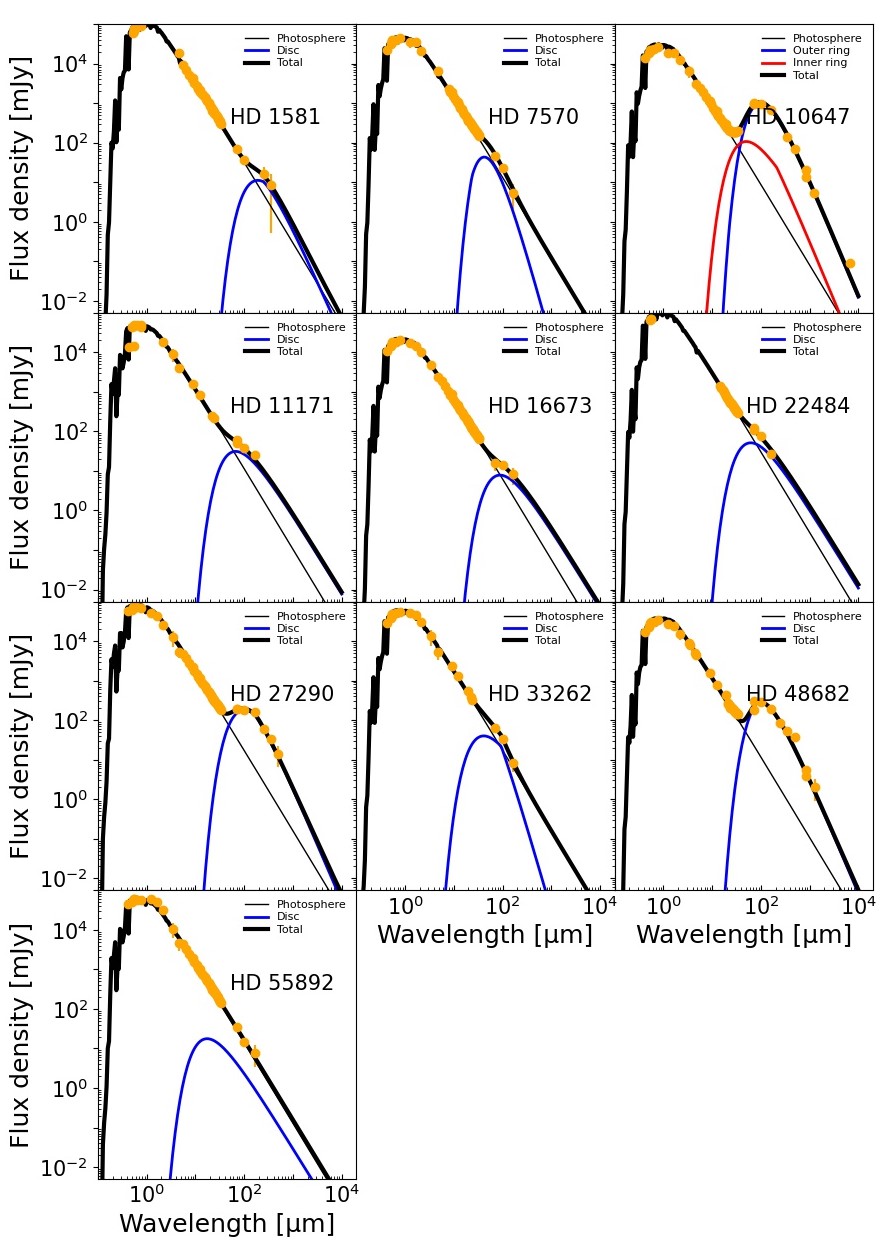}
 \caption{SEDs for the debris discs detected around F~stars in the DEBRIS sample.  
 \label{fig:SEDs_DEBRIS_1}}
\end{figure*}

\begin{figure*}
\includegraphics[width=0.9\textwidth]{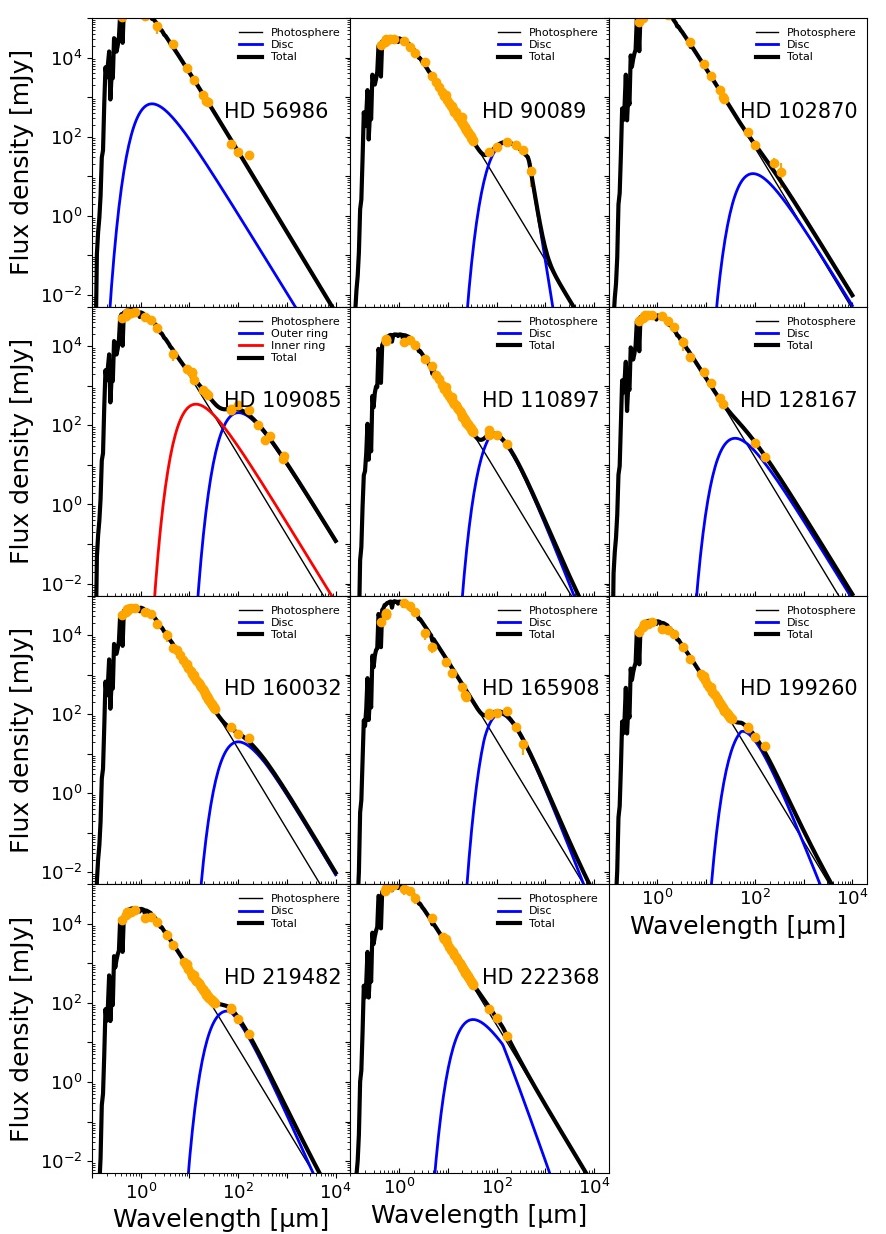}
 \caption{SEDs for the debris discs detected around F~stars in the DEBRIS sample (continued).  
 \label{fig:SEDs_DEBRIS_2}}
\end{figure*}

We analysed the sample of 92 F-type stars in DEBRIS (see \S~\ref{sec:fieldstars}) and found that 21 of them exhibit significant mid-infrared excess. 
We fitted the SEDs of these targets with a modified blackbody model which is described in detail in Section~\ref{sec:IRExcess}. The results are shown in Fig.~\ref{fig:SEDs_DEBRIS_1} and \ref{fig:SEDs_DEBRIS_2}. 

\section{Collisional disc evolution}
\label{sec:collisions}

While we inferred the minimum sizes of dust from SED modelling in \S~\ref{sec:SEDmodelling} we can further constrain the size distribution by 
inferring the minimum size of the planetesimals that must be feeding the collisional cascade, by extrapolating the size distribution of the dust up to the size at which the collisional lifetime is equal to the age of the star, applying the lifetimes calculated using
the analytical collision evolution model introduced by \cite{wyatt-et-al-2007}.

\subsection{Collision model}

The model uses a similar power law size distribution as the SED model following  
\begin{equation}
N_\text{coll}(s) \approx s^{2-3q},
\end{equation}
with $s$ being the radius of a spherical body and $N(s)ds$ the number of bodies in the size range $s$ to $s+ds$.
We note that the parameter $q$ is different from the size distribution index inferred by SED modelling (\S~\ref{sec:SEDmodelling}). They are related by $q_\text{SED} = -(2-3q)$ leading to $q_\text{SED} = 3.5$ and $q = 1.83$ for an ideal collisional cascade \citep{dohnanyi-1969}.

Following eqs.~(12) from \cite{wyatt-et-al-2007} and (22) from \cite{loehne-et-al-2007} we get an equation for the collisional timescale, $t_\text{c}$ as a function of minimum size of the planetesimals necessary to feed to collisional cascade, $s_\text{c}$: 
\begin{equation}
\begin{split}
t_\text{c} = \frac{r^{1/2}\,dr\,i}{(\gamma M_\text{star})^{1/2} f_\text{d} \left(\frac{5}{4}e^2+i^2\right)^{1/2}} \left(\frac{s_\text{c}}{s_\text{blow}}\right)^{3q-5} \\
\left\{\left[X_\text{c}^{5-3q}-1\right]
+ 2\frac{q-5/3}{q-4/3}\left[X_\text{c}^{4-3q}-1\right]
+ \frac{q-5/3}{q-1}\left[X_\text{c}^{3-3q}-1\right]\right\}^{-1}.
\end{split}
\label{eq:timescale}
\end{equation}
The timescale depends on the fractional luminosity, $f_\text{d}$, the blow-out grain size, $s_\text{blow}$, the stellar mass, $M_\text{star}$, the disc radius, $r$, the disc width, $dr$, the eccentricity, $e$, the inclination, $i$, and the parameter $X_c$ which is defined as 
\begin{equation}
X_c = \left[\frac{2Q_\text{D}^*}{v^2_\text{imp}}\right]^{1/3}.
\end{equation}
Here, $Q_\text{D}^*$ the catastrophic disruption threshold and $v_\text{imp}$ the impact velocity of the colliding bodies given as 
$v_\text{imp}~=~\sqrt{\gamma M_\text{star}\, r^{-1}\,(5/4e^2+i^2)}$ with $\gamma$ as gravitational constant. 
In the following section we fix both eccentricity and inclination to a value of 0.1.

\subsection{The catastrophic disruption threshold}

The collisional timescale strongly depends on the catastrophic disruption threshold, $Q_\text{D}^*$ of the planetesimals which is the specific energy necessary to disperse a target \citep[e.g.,][]{benz-asphaug-1999}.  
The parameter can be described by a two-power law function taking into account the material strength, the self-gravity of particles and the impact velocity \citep{stewart-leinhardt-2009}:
\begin{equation}
Q_\text{D}^* = \left[A\left(\frac{s}{1{\rm cm}}\right)^a + B \left(\frac{s}{1{\rm cm}}\right)^b\right]\left(v_\text{imp}\right)^{k}.
\label{eq:QD}
\end{equation}
The parameters $A$, $B$, $a$, $b$ and $k$ are material constants.
We found that $v_\text{imp}$ is on average $\sim0.4$~km/s for the discs in our samples assuming that $e=i=0.1$.
Following \cite{obrien-greenberg-2003}, we can infer the size distribution index, $q_\text{SED}$, from $Q_\text{D}^*$ using the parameter $a$ from eq.~(\ref{eq:QD}): 
\begin{equation}
q_\text{SED} = \frac{7+a/3}{2+a/3}.
\end{equation}
Hence, not only the collisional timescale but also the size distribution of planetesimals depends on $Q_\text{D}^*$.

\begin{figure}
\includegraphics[angle = -90, width=\columnwidth]{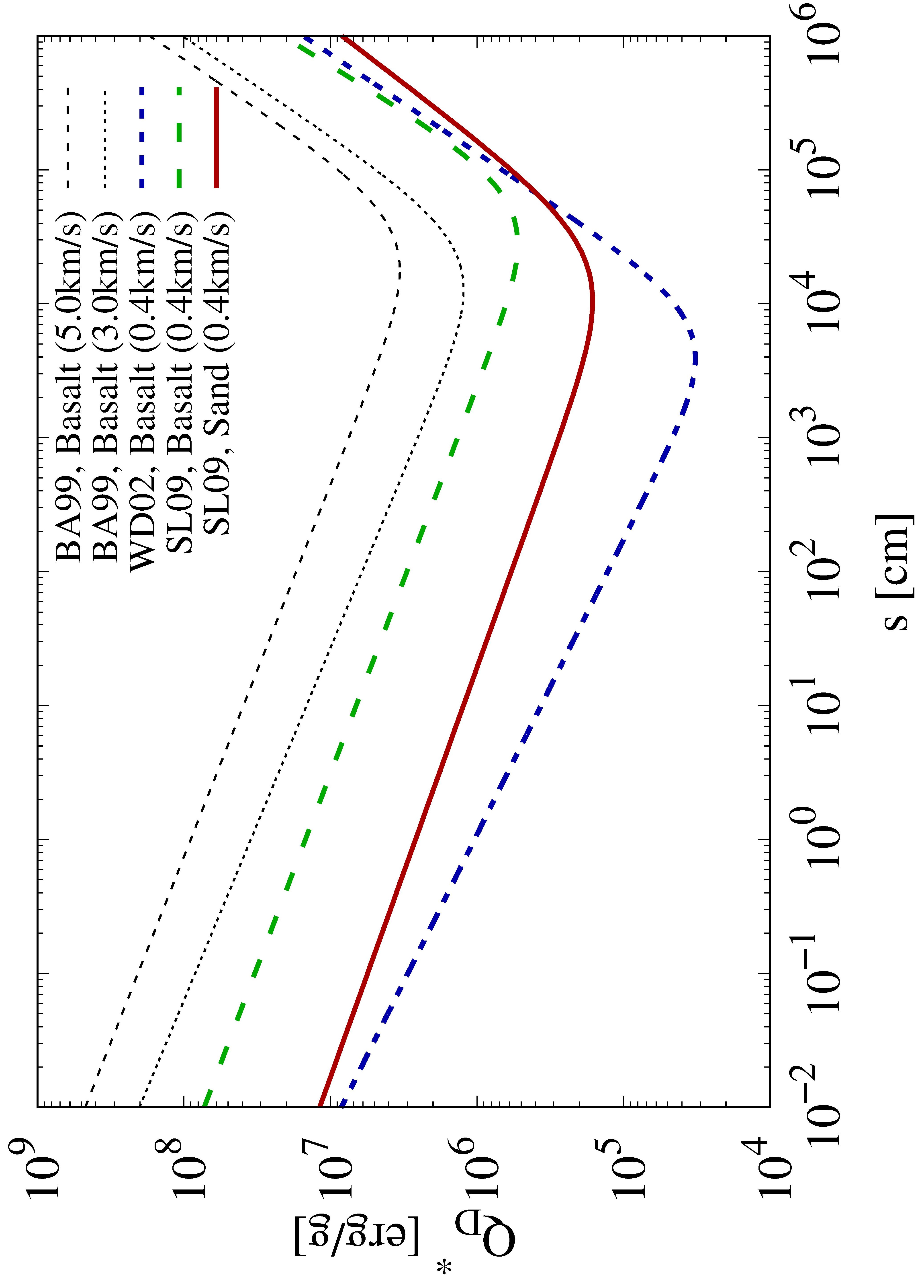}
 \caption[]{\label{fig:QD}
 $Q_\text{D}^*$ as function of size. The thin black dashed and dotted line show values for basalt at 3 and 5~km/s taken from \protect{\cite{benz-asphaug-1999}}. The parameters of the HD~109085 system are assumed. The thick blue dash-dotted line shows the scaling result taken for \protect{\cite{wyatt-dent-2002}} and the thick green dashed line the lab results taken from \protect{\cite{stewart-leinhardt-2009}} for basalt at 0.4~km/s. The thick solid red line shows the result for ``sand'' at 0.4~km/s taken from \protect{\cite{stewart-leinhardt-2009}}.
 }
\end{figure}

Studies of the collisional evolution of debris discs \citep[e.g.,][]{wyatt-dent-2002, schueppler-et-al-2014, loehne-et-al-2017, krivov-et-al-2018, geiler-et-al-2019} often assume the materials ``sand'' \citep[][]{stewart-leinhardt-2009} and basalt \citep{benz-asphaug-1999}.  
In Fig.~\ref{fig:QD}, $Q_\text{D}^*$ is depicted as a function of size for both materials.
Using eq.~(\ref{eq:QD}), we find that the values for basalt colliding at 0.4~km/s lie one order of magnitude below those of \cite{benz-asphaug-1999} colliding at 5~km/s. 

Another approach to infer values of $Q_\text{D}^*$ at the appropriate impact velocity is given by \cite{wyatt-dent-2002} which introduced a scaling method where 
$Q_\text{D}^* \propto v_\text{imp}^{\delta}$. Here, $\delta$ is found by comparing the two impact velocity curves given in \cite{benz-asphaug-1999}.
The method gives values one order of magnitude below those from \cite{stewart-leinhardt-2009} for sizes smaller than $\sim100$~m (strength regime). For larger sizes ($\sim1$~km, gravity regime) the values are comparable to each other. 
The results using sand as material at 0.4~km/s lie between those of basalt from \cite{wyatt-dent-2002} and \cite{stewart-leinhardt-2009} and show a flatter decrease than basalt in the strength regime. The values in the gravity regime are close to those of basalt, but the increase is flatter.

The approaches of \cite{wyatt-dent-2002} and \cite{stewart-leinhardt-2009} to scale $Q_\text{D}^*$ to the impact velocity are both used in the literature. Therefore, we emphasise that $Q_\text{D}^*$ strongly depends on the method applied and shows variations of one order of magnitude even for the same material. Furthermore, $Q_\text{D}^*$ varies for the material chosen.

\subsection{Minimum sizes of planetesimals feeding the cascade}
\label{sec:planetesimalsizes}
We calculate the minimum sizes of the planetesimals feeding the collisional cascade using eq.~(\ref{eq:timescale}) assuming that the collisional timescale, $t_\text{c}$, is similar to the age of the system.

\begin{table*}
\caption{Sizes of planetesimals.
\label{tab:planetesimals}}
\tabcolsep 4pt
\centering
 \begin{tabular}{r|rccccc|cccccc}
       & \multicolumn{6}{c}{System parameters}           & \multicolumn{2}{c}{Basalt (WD02)}   & \multicolumn{2}{c}{Basalt (SL09)}    & \multicolumn{2}{c}{Sand (SL09)} \\
HD     & r    & dr   & $M_\text{star}$ & $s_\text{blow}$ & $f_\text{d}$   & $t_\text{c}$ & $s_\text{c}$ & $M_\text{disc}$  & $s_\text{c}$ & $M_\text{disc}$ & $s_\text{c}$ & $M_\text{disc}$  \\
       & [au] & [au] & [$M_\odot$]     & [$\mu$m ]       &                & [Myr]        & [m]          & [$M_\text{earth}$] & [m]          & [$M_\text{earth}$] & [m]          & [$M_\text{earth}$] \\
	\midrule
 15115 &93    & 21   & 1.37            & 0.91 & $5.3\times10^{-4}$ & 23           &  338  &  23    &  104 &   7.0 &  381 &  26   \\ 
160305 &88    &  4   & 1.13            & 0.67 & $1.5\times10^{-4}$ & 23           &  351  &   6.0  &  115 &   2.0 &  401 &   6.9 \\ 
164249 &63    & 24   & 1.30            & 0.89 & $9.4\times10^{-4}$ & 23           &  514  &  28    &  413 &  23   &  678 &  37   \\ 
181327 &81    & 16   & 1.36            & 1.02 & $4.1\times10^{-3}$ & 23           & 1417  & 562    & 1601 & 634   & 2188 & 867   \\ 
191089 &45    & 16   & 1.36            & 0.98 & $1.6\times10^{-3}$ & 23           & 1105  &  53    & 1310 &  63   & 1703 &  81   \\ 
\midrule
10647  &82    & 49   & 1.12            & 0.48 & $2.9\times10^{-4}$ & 1000         &  960  &  28    & 1006 &  29   & 1405 &  40   \\
109085 &152   & 46   & 1.53            & 1.24 & $1.7\times10^{-5}$ & 1000         &  239  &   1.4  &   19 &   0.11&  213 &   1.2 \\
\end{tabular}

\noindent
{\em Notes:}
The system data for HD~10647 were taken from Lovell et al. (in prep.), the data for HD~109085 from \cite{matra-et-al-2018}. The age estimates for both stars show large uncertainties so that we fix the age to 1000~Myr for simplicity reasons. ``Basalt (WD02)'' refers to the scaling method of $Q_\text{D}^*$ used in \cite{wyatt-dent-2002} while ``Basalt (SL09)'' assumes the velocity dependence found in \cite{stewart-leinhardt-2009}. ``Sand (SL09)'' refers to the weak rock material introduced in \cite{stewart-leinhardt-2009}.   
\end{table*}

Tab.~\ref{tab:planetesimals} lists the planetesimal sizes and the corresponding minimum disc masses (since the size distribution must extend up to these sizes, and could extend further) assuming the two different approaches to scale $Q_\text{D}^*$ to the impact velocity of the colliding bodies as well as the two materials basalt and sand.
We added the discs HD~10647 and HD~109085 from the DEBRIS sample to compare the planetesimal sizes in systems of different age. Both discs were spatially resolved with ALMA.

We find that the smallest planetesimals feeding the collisional cascade show sizes from several metres up to $\sim2$ kilometres independently of the age of the system. For some discs this is somewhat smaller than assumed by former studies which found sizes around kilometres \citep[e.g.,][]{wyatt-dent-2002, marino-et-al-2017, krivov-et-al-2018}. This is also smaller than the predicted sizes of hundreds of kilometres from planetesimal formation scenarios \citep[e.g.,][]{klahr-et-al-2020} and might indicate a lack of those large planetesimals as was proposed by \cite{krivov-et-al-2020}. 

However, we find that the discs analysed show variations of sizes of one order of magnitude for all materials. 
Applying eq.~(\ref{eq:QD}) the maximum sizes using basalt are smaller than those using sand. 
While for large planetesimals in the gravity regime the differences between the materials are small they become more pronounced for metre-sized planetesimals close to the strength regime similar to the trend of $Q_\text{D}^*$ shown in Fig.~\ref{fig:QD}.
Considering the two scaling methods we find a comparable trend -- the method chosen to infer $Q_\text{D}^*$ becomes more important for smaller sizes. 

The large variation in sizes leads to different disc masses depending on the material applied. Again, discs for which the planetesimals feeding the dust belt are only required to be metre in size are more sensitive to the material and method used. The masses vary between 2$M_\oplus$ (HD~160305) and 900$M_\oplus$ (HD~181327). 

Studies of planetesimal formation \citep[e.g.,][]{klahr-et-al-2020} found that typical planetesimal sizes tend to decrease with increasing distance to the star and with the time of the formation of planetesimals. 
While an early formation might lead to sizes of $\sim100$~km, planetesimals formed at a later stage tend to be as small as $\sim10$~km \citep[e.g.,][]{stern-et-al-2019}. This is still somewhat larger than the sizes of $\sim1$~km we infer for our discs, but we note that our estimated sizes are minimum sizes necessary to feed the collisional cascade.


\bsp	
\label{lastpage}
\end{document}